\documentclass{article}
\usepackage{amsthm}
\usepackage{amsfonts}
\usepackage{amsmath}
\usepackage[all]{xy}

\numberwithin{equation}{section}
\begin{document}
\newcommand{\Z}{\mathbb{Z}}
\newcommand{\N}{\mathbb{N}}
\newcommand{\R}{\mathbb{R}}
\newcommand{\C}{\mathbb{C}}
\newcommand{\HH}{\mathbb{H}}
\newcommand{\Hom}{\mathrm{Hom}}
\newcommand{\End}{\mathrm{End}}
\newcommand{\sgn}{\mathrm{sgn}}
\newcommand{\rme}{\mathrm{e}}
\newcommand{\rmi}{\mathrm{i}}
\newcommand{\rmd}{\mathrm{d}}

\newtheorem{corollary}{Corollary}%[section]

\title{Clifford modules and symmetries of topological insulators}
\author{Gilles Abramovici and Pavel Kalugin\\ Laboratoire de Physique des Solides, \\Univ. Paris-Sud, CNRS, UMR 8502,\\
F-91405 Orsay Cedex, France\\
{\tt abramovici@lps.u-psud.fr}}
\maketitle
\begin{abstract}
We complete the classification of symmetry constraints on gapped quadratic fermion hamiltonians proposed by Kitaev. The symmetry group is supposed compact and can include arbitrary unitary or antiunitary operators in the Fock space that conserve the algebra of quadratic observables.
We analyze the multiplicity spaces of {\em real} irreducible representations of unitary symmetries in the Nambu space. The joint action of intertwining operators and antiunitary symmetries provides these spaces with the structure of Clifford module: we prove a one-to-one correspondence between the ten Altland-Zirnbauer symmetry classes of fermion systems and the ten Morita equivalence classes of real and complex Clifford algebras. The antiunitary operators, which occur in seven classes, are projectively represented in the Nambu space by unitary ``chiral symmetries''. The space of gapped symmetric hamiltonians is homotopically equivalent to the product of classifying spaces indexed by the dual object of the group of unitary symmetries.
\end{abstract}
%\pacs{03.65.Fd, 03.65.Vf, 73.43.Nq, 74.90.+n}
%\keywords{Topological insulators, Clifford modules}
\section{Introduction}\label{sec:introduction}
The purpose of this paper is to complete the classification of topological insulators and superconductors proposed in \cite{kitaev}. It was conjectured in \cite{schnyder} that the topological phases fall into one of ten possible symmetry classes, corresponding to ten symmetry classes of fermionic hamiltonians \cite{hhz, zirnbauer2010}. For a given spatial dimension topological phases may occur only in five of the ten classes. As the dimension varies, the topological phases are permuted, exhibiting two-fold and eight-fold periodicity. Kitaev \cite{kitaev} has explained this behavior as a manifestation of Bott periodicity in real or complex K-theory. Technically, the link with K-theory originates from the association of gaps in the spectra of quadratic fermionic hamiltonians with extensions of Clifford modules. Until now these modules have been constructed explicitly for 3 out of 10 classes only \cite{kitaev} (mention should also be made of the work \cite{stone}, where a Clifford module structure is suggested for all ten symmetry classes; see section \ref{sec:discussion} for further discussion). We complete this classification and associate the remaining 7 classes to the corresponding Clifford algebras. In addition, we extend and simplify the classification of Fermionic hamiltonians proposed in \cite{hhz, zirnbauer2010}.
\par

This article is organized as follows. In section \ref{sec:statement} we
formulate the problem and state the main result. Section \ref{sec:nambu} deals
with symmetry constraints imposed on the system. We show there how the action of
unitary symmetries can be transferred from the Fock space to the Nambu space. In
section \ref{sec:isotypic} we introduce a version of the canonical isotypic
decomposition customized for the representations in real vector spaces. In
section \ref{sec:antiunitary} we deal with antiunitary symmetries and describe
their action on the isotypic components. Section \ref{sec:gaps} is dedicated to
the main result of the paper: we show that the multiplicity space of each
isotypic component of the Nambu space has the structure of a Clifford module.
This allows one to apply the arguments presented in \cite{kitaev} and define the
homotopy type of the space of gapped symmetric hamiltonians in terms of
classifying spaces of real or complex K-theory. Section \ref{sec:examples} is
devoted to the study of two examples. A final discussion is included in section
\ref{sec:discussion}.

\section{Classification of gapped hamiltonians}\label{sec:statement}
Roughly speaking, the goal of the classification of topological phases consists in answering the following question:

\begin{quote}
Let $H_1$ and $H_2$ be quadratic hamiltonians of the form:
$$
\sum_{i,j} \left( A_{ij} a^\dag_i a_j + B_{ij} a_i a_i + B_{ij}^* a^\dag_j a^\dag_i \right),
$$
where $a_i^\dag$, $a_i$ are the fermionic creation and annihilation operators and the matrices satisfy the conditions $A_{ij}=A_{ji}^*$, $B_{ij}=-B_{ji}$. Suppose that both $H_1$ and $H_2$ are gapped. Is it possible to transform $H_1$ to $H_2$ continuously while preserving the gap?
\end{quote}
Actually, the question above is loosely stated. First, it omits to mention the symmetry constraints imposed on the system. The word ``system'' here is misnamed since symmetries always characterize families of systems and not individual ones. In this regard fixing the symmetry group is always a matter of voluntary choice\footnote{The choice is often guided by the apparent symmetry of the classical system. The symmetry group $G_{\rm c}$ of the latter is then assumed to act on the space $V$ of single-particle states by a projective representation, which can be lifted to a linear unitary representation of a central extension of $G_c$ \cite{projective_rep}. Other symmetries may be hidden, as for instance the $SO(4)$ symmetry in hydrogen-like atoms \cite{fock} or that of isospin in graphene monolayers \cite{isospin}.}. We shall thus follow \cite{hhz} on this point and treat the symmetries as externally imposed conditions. Specifically, we fix a compact group $G_{\rm F}$ and its action on the Fock space by unitary or antiunitary operators. Only hamiltonians commuting with the action of $G_{\rm F}$ are taken into consideration. This constraint applies to all hamiltonians along the entire path of continuous transformation between $H_1$ and $H_2$ as well. Namely, the path $H(\tau)$ parameterized by $\tau$, with $H(0)=H_1$ and $H(1)=H_2$, is allowed only when $H(\tau)$ commutes with the action of $G_{\rm F}$ for all $\tau\in [0,1]$. For convenience, we shall further restrict the class of allowed hamiltonians by imposing the condition of zero trace:
\begin{equation}
\label{traceless_quadratic}
H=\sum_{i,j} \left( \frac{1}{2}A_{ij} \left(a^\dag_i a_j - a_j a^\dag_i \right) + B_{ij} a_i a_i + B_{ij}^* a^\dag_j a^\dag_i \right).
\end{equation}
Eventually, the gap will always refer to that in the single-particle spectrum of $H$, that is the spectrum of noninteracting fermions, obtained by the appropriate Bogoliubov transformation.
\par
The question above is also too restrictive. Considering, for instance, the adiabatic pumping problem, one will study hamiltonians which depend on external controllable parameters (e.g. the gate voltage) \cite{thouless}. The question of interest is then the existence of closed loops in the space of gapped Hamiltonians which cannot be contracted without closing the gap. Thus, the deformation of $H$ must preserve the loop. In this case the connectedness of the space $\cal S$ of allowed hamiltonians may not suffice to describe the topological phase. The right question is that of the homotopy equivalence class of $\cal S$.
\par
This question is eventually answered in section \ref{sec:gaps}. Formula (\ref{S_product}) describes the structure of a deformation retract of ${\cal S}$ as a cartesian product of some specific classifying spaces, indexed by the classes of irreducible representations of the unitary symmetries in the real part of the Nambu space. Each classifying space belongs to one of the ten symmetry classes, given by formulas (\ref{sum_to_simple}), (\ref{simple_to_sum}) and (\ref{simple_to_simple}).
\section{Nambu space and its symmetries}\label{sec:nambu}
In this section we first recall the construction of Nambu space representation of fermionic systems with quadratic hamiltonians. We then describe the class of unitary symmetries in the Fock space that are compatible with the quadratic observables, and define the representation of these operators in the Nambu space. Finally, we define the real structure in the Nambu space and use it to characterize the groups of allowed symmetries in both spaces.
\subsection{Nambu space representation}
Let $V$ denote the $N\mbox{-dimensional}$\footnote{We assume throughout the paper that the number of fermionic states is finite. However, whenever possible the provided proofs apply to the case of infinite dimension. See also the discussion in section \ref{sec:discussion}.} complex Hilbert space of single-particle states and $\langle \, ,\rangle$ be the corresponding sesquilinear form (antilinear in the first and linear in the second argument). The $2^N\mbox{-dimensional}$ Fock space of fermionic system is defined as the exterior algebra over $V$:
$$
\wedge V = \bigoplus_{n=0}^N \wedge^n V,
$$
where $\wedge^n V$ represents the space of $n\mbox{-particle}$ states (with $\wedge^0 V= \C$ for the vacuum state). The creation operator $a^\dag_v$ of one particle in the state $v \in V$ is represented in $\wedge V$ by the exterior multiplication by $v$, and the corresponding annihilation operator $a_v$ by the contraction with the linear form $\langle v, \cdot \rangle$ (cf. \cite{hhz} for more details).
\par
In what follows we shall concentrate on the special situation when the hamiltonian and all physical quantities of interest belong to the space ${\cal Q}$ of traceless quadratic operators of the form (\ref{traceless_quadratic}). ${\cal Q}$ is closed with respect to the operation
$$
[X, Y] = \rmi(XY-YX),
$$
which provides it with the structure of a real Lie algebra. Since $\cal Q$ is much smaller than the algebra of generic hermitian operators, one can expect that it admits a faithful representation in a space smaller than $\wedge V$. To prove the existence of such representation, let us construct it explicitly. Let $V^*$ denote the dual space of $V$. Consider the mapping $\alpha$ from the complex linear space
$$
W=V \oplus V^*
$$
to the space of operators in $\wedge V$, defined by the formula
\begin{equation}
\label{mu}
\alpha\left(v_1 + \langle v_2, \cdot \rangle \right) = a^\dag_{v_1}+a_{v_2}.
\end{equation}
Let $X \in {\cal Q}$. Then for any $w \in W$ there exists $w' \in W$ such that
\begin{equation}
\label{comm}
[X, \alpha(w)]=\alpha(w').
\end{equation}
Since $\alpha$ is $\C\mbox{-linear}$, $w \mapsto w'$ is linear, that is there exists a complex linear operator $X_{\rm N}: W \to W$ such that
$$
X_{\rm N}(w)=w'.
$$
The mapping $X \mapsto X_{\rm N}$ is defined for every $X \in {\cal Q}$ and is clearly a representation of the Lie algebra ${\cal Q}$ in $W$. The $2N\mbox{-dimensional}$ complex space $W$ is commonly referred to as the ``particle-hole'' or Nambu space.
\par
The Nambu space representation makes it possible to use the first quantization for systems with non-conserved particles, such as superconductors. The evolution of the system is described by the ``vector of state'' $w \in W$ obeying the equation
\begin{equation}
\label{evolution}
\dot w = \rmi H_{\rm N} w,
\end{equation}
where $H_{\rm N}$ stands for the representation of the hamiltonian in the Nambu space. The state $w=v_1 + \langle v_2, \cdot \rangle$ can be interpreted as a superposition of a particle in the state $v_1$ and a hole in the state $v_2$.
\subsection{${\cal Q}\mbox{-compatible}$ unitary symmetries in Fock and Nambu spaces}
Let us consider now the symmetries of the system. They are described by a group $G_{\rm F}$ acting on $\wedge V$ by unitary or antiunitary operators. Contrary to \cite{hhz, zirnbauer2010} we shall not require that the action of $G_{\rm F}$ on $\wedge V$ originates from the symmetries of the single particle space $V$. Instead, we impose a weaker condition of compatibility with algebra ${\cal Q}$:
\begin{quote}
A linear or antilinear invertible operator $u$ in the space $\wedge V$ is said to be ${\cal Q}\mbox{-compatible}$ if $u{\cal Q}u^{-1}={\cal Q}$.
\end{quote}
Let us describe the set of ${\cal Q}\mbox{-compatible}$ symmetries in details. We shall consider here the subgroup $G_{{\rm F}0}\subset G_{\rm F}$ of unitary operators only, leaving antiunitary ones for section \ref{sec:antiunitary}. Let $\cal K$ be the subgroup of $\mathrm{U}(\wedge V)$ obtained by exponentiation of ${\cal Q}$:
\begin{equation}
\label{spin}
{\cal K}=\left\{\exp\left(\rmi X_{\rm F}\right)\mid X_{\rm F} \in {\cal Q}\right\}.
\end{equation}
By construction, the full set of ${\cal Q}\mbox{-compatible}$ unitary operators is the normalizer of ${\cal K}$ in $\mathrm{U}(\wedge V)$. This subgroup, denoted by $N_{\mathrm{U}(\wedge V)}\left({\cal K}\right)$, acts on ${\cal K}$ by conjugation. For some elements of $N_{\mathrm{U}(\wedge V)}\left({\cal K}\right)$ this action is an inner automorphism of ${\cal K}$. These elements form a subgroup of $N_{\mathrm{U}(\wedge V)}\left({\cal K}\right)$, which we shall denote ${\cal K}_c$. Since the only unitary operators commuting with all elements of ${\cal K}$ are scalars, this group consists of the elements of the following form:
\begin{equation}
\label{spin_c}
{\cal K}_c=\left\{\exp\left(\rmi \left( X_{\rm F}+c \right) \right)\mid X_{\rm F} \in {\cal Q},\; c \in \R\right\}.
\end{equation}
Clearly, ${\cal K}_c$ is a normal subgroup in $N_{\mathrm{U}(\wedge V)}\left({\cal K}\right)$ which cosets correspond to outer automorphisms of ${\cal K}$:
$$
N_{\mathrm{U}(\wedge V)}\left({\cal K}\right) / {\cal K}_c \subset \mathrm{Out}({\cal K}).
$$
On the other hand, since ${\cal K}$ is obtained by exponentiation of ${\cal Q}$
within the Clifford algebra of fermionic field operators, it is isomorphic to
$\mathrm{Spin}(2N)$ \cite{heat_kernel}. For $N>4$, one has $\mathrm{Out}\left(\mathrm{Spin}(2N)\right)=\Z_2$, that is there is only one non-trivial outer automorphism \cite{wilson}. Therefore, one element of $N_{\mathrm{U}(\wedge V)}\left({\cal K}\right)$ not belonging to ${\cal K}_c$ will by itself fully characterize the set of ${\cal Q}\mbox{-compatible}$ unitary operators. Let us construct this element explicitly. Let $w_0 \in W$ be given by
$$
w_0=e_0+\langle e_0, \cdot \rangle,
$$
where $e_0 \in V$ is a vector of unit norm. One can readily check that
\begin{equation}
\label{mu_Q_compatible}
\alpha(w_0)\,{\cal Q}\,\alpha(w_0) = {\cal Q},
\end{equation}
where $\alpha$ is defined in (\ref{mu}). As $\alpha(w_0)^2=1$, the operator $\alpha(w_0)$ is unitary and, because of (\ref{mu_Q_compatible}) it is also ${\cal Q}\mbox{-compatible}$. On the other hand, $\alpha(w_0) \notin {\cal K}_c$, since it changes the parity of the number of particles. Therefore, we obtain the following characterization:
\begin{quote}
\em{
All parity-conserving ${\cal Q}\mbox{-compatible}$ unitary operators in $\wedge V$ have the form
\begin{equation}
\label{parity_conserving}
\exp\left(\rmi \left( X_{\rm F}+c \right) \right),
\end{equation}
where $X_{\rm F}\in {\cal Q}$ and $c\in \R$. All ${\cal Q}\mbox{-compatible}$ operators that change the parity of the number of particles have the form
\begin{equation}
\label{parity_changing}
(a^\dag_{e_0}+a_{e_0})\exp\left(\rmi \left( X_{\rm F}+c \right) \right),
\end{equation}
where $e_0$ is a state of unit norm.}
\end{quote}
\par
We shall define the representation of a ${\cal Q}\mbox{-compatible}$ unitary operator $U_{\rm F}$ in the Nambu space by the requirement that its action by conjugation on $X_{\rm N}$ matches the action of $U_{\rm F}$ on $X_{\rm F}$ for any $X_{\rm F}\in{\cal Q}$:
\begin{equation}
\label{fock-nambu}
\left(U_{\rm F} X_{\rm F} U_{\rm F}^{-1}\right)_{\rm N} = U_{\rm N} X_{\rm N} U_{\rm N}^{-1},
\end{equation}
where $X_{\rm N}$ is the Nambu representation of $X_{\rm F}$. For
parity-conserving operators con\-dition (\ref{fock-nambu}) is satisfied by exponentiation of the action of the Lie algebra ${\cal Q}$ in~$W$:
\begin{equation}
\label{spin_nambu}
\left(\exp\left(\rmi \left(X_{\rm F}+c\right)\right)\right)_{\rm N}=\exp\left(i X_{\rm N}\right).
\end{equation}
It remains to define the Nambu representation of $\alpha(w_0)$. A little algebra shows that for any $w \in W$ there exists $w' \in W$ such that
\begin{equation}
\label{adjoint}
\alpha(w_0)\alpha(w)\alpha(w_0)^{-1}=\alpha(w')
\end{equation}
(recall that $\alpha(w_0)^{-1}=\alpha(w_0)$). Let us define the complex linear map $\rho_0: W \to W$ by the formula
\begin{equation}
\label{twisted}
\rho_0(w)=-w'.
\end{equation}
We set $\rho_0$ as the Nambu representation of $\alpha(w_0)$:
\begin{equation}
\label{mu_nambu}
\alpha(w_0)_{\rm N}=\rho_0.
\end{equation}
One can check that this definition satisfies condition (\ref{fock-nambu}):
\begin{equation}
\label{alpha-rho}
\left(\alpha(w_0) X_{\rm F} \alpha(w_0)^{-1}\right)_{\rm N} = \rho_0 X_{\rm N} \rho_0^{-1}.
\end{equation}
Exponentiation of (\ref{alpha-rho}) shows that formulas (\ref{spin_nambu}) and (\ref{mu_nambu}) define a group homomorphism and therefore establish the Nambu space representation of ${\cal Q}\mbox{-compatible}$ unitary operators. Note also that, if one considers operators (\ref{parity_conserving}) and (\ref{parity_changing}) with $c=0$ only, the above construction is equivalent to the twisted adjoint representation of group $\mathrm{Pin}(2N)$, defined in \cite{abs}.
\subsection{The real structure of the Nambu space}
The geometric meaning of the above construction is best understood if one takes into account the {\em natural real structure} of the complex Nambu space $W$. This structure is given by the antilinear involution $C: W \to W$, which is defined on $V$ by the formula
\begin{equation}
\label{C}
C(v)=\langle v, \cdot \rangle \in V^*,
\end{equation}
and on $V^*$ by the condition $C^2=1$. Since the anticommutator of operators (\ref{mu}) reads as
$$
\left\{\alpha(w_1), \alpha(w_2)\right\}=2\left\langle C(w_1), w_2 \right\rangle,
$$
$C$ is invariant with respect to any symmetry operation. Denote by $W_\R$ the set of $C\mbox{-invariant}$ vectors in $W$:
\begin{equation}
\label{W_R}
W_\R = \left\{w\in W \mid Cw=w \right\}
\end{equation}
$W_\R$ is a $2N\mbox{-dimensional}$ real vector space\footnote{The operators $\alpha(w)$ with $w \in W_\R$ are usually referred to as Majorana fermions.} with an Euclidean structure inherited from the sesquilinear form in $W$. With respect to this structure operators (\ref{traceless_quadratic}) are imaginary antisymmetric. Hence, the parity-conserving ${\cal Q}\mbox{-compatible}$ unitary operators (\ref{spin_nambu}) correspond to proper orthogonal transformations of $W_\R$. On the other hand, $\rho_0$ defined by formulas (\ref{adjoint}) and (\ref{twisted}) acts on $W_\R$ by reflection in the hyperplane orthogonal to $w_0$. Therefore, any ${\cal Q}\mbox{-compatible}$ unitary operator is represented by an isometry of $W_\R$.
\par
We are now ready to describe in details the relation between the Fock and Nambu representations of ${\cal Q}\mbox{-compatible}$ unitary symmetries. A parity-conserving operator (\ref{spin_c}) is a product of an element of ${\cal K}=\mathrm{Spin}(2N)$ and a complex unitary scalar. Therefore, they form a subgroup of $\mathrm{U}(\wedge V)$ isomorphic to
$$
\mathrm{Spin}^\C(2N)=\left(\mathrm{Spin}(2N) \times U(1)\right)/\{+1,-1\}.
$$
The representation of (\ref{spin_c}) in $W_\R$ corresponds to a homomorphism $\mathrm{Spin}^\C(2N) \to \mathrm{SO}(W_\R)$ (we will use this notation instead of that of abstract group $\mathrm{SO}(2N)$). On the other hand, the parity-changing operators are represented by improper orthogonal transformations of $W_\R$. Therefore, the full group of ${\cal Q}\mbox{-compatible}$ unitary operators is isomorphic to the group $\mathrm{Pin}^\C(2N)$ as shown in the following commutative diagram
\begin{equation}
\label{pin}
\xymatrix{\mathrm{Spin}^\C(2N)\ar[r]\ar[d] & \mathrm{Pin}^\C(2N)\ar[r]\ar[d]& \Z_2 \ar@{=}[d] \\
\mathrm{SO}(W_\R)\ar[r] & \mathrm{O}(W_\R)\ar[r]& \Z_2}
\end{equation}
The rows of the diagram are short exact. The groups in the top row act in the
Fock space, and those in the bottom row correspond to the Nambu representation. The groups in the left column represent the parity-conserving operators, while the right column corresponds to the action on the parity of the number of particles.
\par
Diagram (\ref{pin}) shows that the Nambu representation of unitary symmetries is not faithful. Its kernel $\mathrm{U}(1) \subset \mathrm{Pin}^\C(2N)$ acts on the Fock space by multiplication by complex unitary scalars. Since the scalars commute with the hamiltonian, one can assume that they always belong to $G_{{\rm F}0}$. Let us define the group $G_0$ as the quotient $G_{{\rm F}0}/\mathrm{U}(1)$. The following commutative diagram shows that $G_0$ acts effectively on the Nambu space $W_\R$:
\begin{equation}
\label{GF0_G0}
\xymatrix{\mathrm{U}(1)\ar[r]\ar@{=}[d] & G_{{\rm F}0}\ar[r]\ar@{^(->}[d]& G_0 \ar[d] \\
\mathrm{U}(1)\ar[r] & \mathrm{Pin}^\C(2N)\ar[r]& \mathrm{O}(W_\R)}
\end{equation}
Therefore, $G_0$ represents faithfully the unitary symmetry constraints on the hamiltonian.
\section{Isotypic decomposition of the Nambu space}\label{sec:isotypic}
In this section we follow the method of \cite{hhz}, and decompose the Nambu space into a sum of blocks corresponding to different equivalence classes of the irreducible representations of $G_0$. The main distinction of our approach from that of \cite{hhz} consists in using the real representations instead of the complex ones. This allows us to take advantage of the natural real structure (\ref{C}) of the Nambu space and to simplify the calculations considerably. Since the isotypic decomposition of real vector spaces is little used, it is worthwhile to consider the related algebraic structures in details.
\par
Let us first recall the fact (see e.g. \cite{goodman_isotypic}) that {\em complex unitary} representations of a compact group admit a unique isotypic decomposition. In the case of the Nambu space this decomposition reads
\begin{equation}
\label{isotypic_c}
W=\bigoplus_{\lambda \in \hat G_0} {\cal W}_\lambda^\C,
\end{equation}
where the summation is performed over all irreducible unitary representations of $G_0$ and indexed by $\hat G_0$, the unitary dual of $G_0$. The set $\hat G_0$ is in general a noncommutative space \cite{connes}, but for compact groups it is a countable set of isolated points with an ordinary discrete topology. Each block ${\cal W_\lambda^\C}$ contains the irreducible representations of type $\lambda$ only, and there exists a canonical isomorphism
\begin{equation}
\label{canonical}
{\cal W}_\lambda^\C={\cal R}_\lambda^\C \otimes_\C \Hom_{G_0}\left( {\cal R}_\lambda^\C, W \right).
\end{equation}
Here ${\cal R}_\lambda^\C$ is a given irreducible representation of type $\lambda$, and the space of $G_0\mbox{-equivariant}$ homomorphisms $\Hom_{G_0}\left( {\cal R}_\lambda^\C, W \right)$ is commonly referred to as the multiplicity space of ${\cal R}_\lambda^\C$.
\par
The decomposition (\ref{isotypic_c}) does not allow one to decouple the action
of $H_{\rm N}$ and $G_0$ on $W$. Indeed, when $\lambda$ and $\lambda^*$ are
conjugate representation types, the action of $H_{\rm N}$ on ${\cal
W}_\lambda^\C$ and ${\cal W}_{\lambda^*}^\C$ are mutually dependent.
As a consequently, one has to work in ${\cal W}_\lambda^\C\oplus {\cal W}_{\lambda^*}^\C$ which leads to cumbersome expressions. This difficulty is the manifestation of the invariance of the real structure (\ref{C}) with respect to ${\cal Q}\mbox{-compatible}$ unitary symmetries and to the time evolution (\ref{evolution}). One can eliminate this constraint by working within the $C\mbox{-invariant}$ real subspace $W_\R$ defined in (\ref{W_R}). This choice has another benefit, which makes itself clear when the antiunitary symmetries are taken into consideration (this was also Dyson's argument in \cite{dyson}). As we shall see in section \ref{sec:antiunitary}, the antiunitary operators in the Nambu space representation act on $W_\R$ by ordinary orthogonal transformations.
\par
So our goal is to construct a canonical isotypic decomposition of the real vector space $W_\R$ in a way similar to (\ref{isotypic_c}) and (\ref{canonical}). Let us denote by $\hat G_0^{\rm r}$ the set of equivalence classes of {\em real} irreducible representations of $G_0$. Since the action of the complex conjugation on $\hat G_0$ transposes conjugate representation types $\lambda$ and $\lambda^*$ when they are not equal, $\hat G_0^{\rm r}$ is the factor of $\hat G_0$ by the orbits of this action. Let ${\cal R}_\lambda$ be a real irreducible representation for some $\lambda \in \hat G_0^{\rm r}$. One can define $E_\lambda$ the multiplicity space of ${\cal R}_\lambda$ in $W_\R$:
\begin{equation}
\label{g_equivar}
E_\lambda = \Hom_{G_0}\left( {\cal R}_\lambda, W_\R \right).
\end{equation}
$E_\lambda$ is immediately an $\R\mbox{-module}$, but can also be equipped with an a richer algebraic structure. Indeed, let us consider the set of the intertwining operators of ${\cal R}_\lambda$:
\begin{equation}
\label{intertwining}
F_\lambda=\Hom_{G_0}\left( {\cal R}_\lambda, {\cal R}_\lambda \right).
\end{equation}
By virtue of Schur lemma $F_\lambda$ is a division algebra over $\R$, which leaves us with only three possibilities: the real numbers $F_\lambda=\R$, the complex numbers $F_\lambda=\C$ and the quaternions $F_\lambda=\HH$ (in Dyson's terms, the corresponding unitary representations in $W$ are of potentially real, complex or pseudo-real types \cite{wigner}). Since $G_0$ acts on ${\cal R}_\lambda$ by left multiplication, ${\cal R}_\lambda$ is naturally a left module over the group algebra $\R G_0$ (and so is $W_\R$). The action of $F_\lambda$ commutes with that of $\R G_0$ and hence can be conveniently written as a multiplication on the right, providing ${\cal R}_\lambda$ with the structure of a $\left(\R G_0, F_\lambda\right)\mbox{-bimodule}$. The $G_0\mbox{-equivariance}$ of homomorphisms in (\ref{g_equivar}) absorbs the left $\R G_0\mbox{-module}$ structure of both arguments. The resulting space $E_\lambda$ is thus equipped with the left action of $F_\lambda$, which is defined by the formula
\begin{equation}
\label{f_action}
(fe)(r)=e(rf),
\end{equation}
where $e \in E_\lambda$, $f \in F_\lambda$ and $r \in {\cal R}_\lambda$.
\par
We are now ready to describe the isotypic decomposition of $W_\R$. The following corollary generalizes the standard primary decomposition \cite{goodman_isotypic} to the case of representations over a non algebraically closed field such as $\R$ (the proof is given in the Appendix):
\begin{corollary}
\label{cor1}
Let $\epsilon_\lambda$ be the tautological evaluation map from the cartesian product ${\cal R}_\lambda \times E_\lambda$ to $W_\R$:
\begin{eqnarray}
\label{evaluation}
\epsilon_\lambda: {\cal R}_\lambda \times E_\lambda \to W_\R \\
\epsilon_\lambda(r, e) = e(r) \nonumber,
\end{eqnarray}
where $e \in E_\lambda$ and $r \in {\cal R}_\lambda$. The map (\ref{evaluation}) factors through the tensor product ${\cal R}_\lambda \otimes_{F_\lambda} E_\lambda$ yielding
\begin{equation}
\label{iota_lambda}
\iota_\lambda: {\cal R}_\lambda \otimes_{F_\lambda} E_\lambda \to W_\R.
\end{equation}
The space $W_\R$ admits an isotypic composition
\begin{equation}
\label{isotypic}
W_\R=\bigoplus_{\lambda\in\hat G_0^{\rm r}} {\cal W}_\lambda
\end{equation}
where the isotypic component ${\cal W}_\lambda$ is defined as the union of all $G_0\mbox{-submodules}$ of $W_\R$ equivalent to ${\cal R}_\lambda$. The map (\ref{iota_lambda}) establishes a canonical isomorphism
between ${\cal R}_\lambda \otimes_{F_\lambda} E_\lambda$ and ${\cal W}_\lambda$:
\begin{equation}
\label{lambda_block}
{\cal W}_\lambda={\cal R}_\lambda \otimes_{F_\lambda} E_\lambda .
\end{equation}
\end{corollary}
It should be pointed out that the simplicity of formula (\ref{lambda_block}) is deceptive since the tensor product in it is taken over $F_\lambda$ and not over the background field $\R$. Specifically, this means that care must be taken to insure that expressions like $r \otimes_{F_\lambda} e$ are well defined, that is they are middle-linear with respect to the multiplication by scalars of $F_\lambda$.
\par
Let us now consider the action of the symmetry group and of the hamiltonian on isotypic components (\ref{lambda_block}). By definition of the evaluation map, $G_0$ acts trivially on the factor $E_\lambda$ in the tensor product (\ref{lambda_block}). Therefore, the Nambu representation of $g\in G_0$ decomposes as
\begin{equation}
\label{g_action}
g_{\rm N}=\bigoplus_{\lambda \in \hat G_0^{\rm r}} g \otimes_{F_\lambda} 1
\end{equation}
(we use the same symbol $g$ to denote both an element of $G_0$ and its representation in ${\cal R}_\lambda$; the exact meaning of $g$ will always be clear from the context). On the other hand, the hamiltonian $H_{\rm N}$ (as well as any other $G_0\mbox{-equivariant}$ operator) can not mix different blocks in the decomposition (\ref{isotypic}). Within any block of the type (\ref{lambda_block}) the action of $\rmi H_{\rm N}$ can be decomposed as
\begin{equation}
\label{h_tensor}
\left. \rmi H_{\rm N} \right|_{{\cal R}_\lambda \otimes_{F_\lambda} E_\lambda} = \sum_j \gamma_{\lambda,j}\otimes_{F_\lambda} \zeta_{\lambda,j}
\end{equation}
Since $\rmi H_{\rm N}$ commutes with any action of the type (\ref{g_action}), $\gamma_{\lambda,j}$ is an intertwining operator of ${\cal R}_\lambda$. Therefore, the action of $\gamma_{\lambda,j}$ corresponds to the right multiplication by some $f_{\lambda,j} \in F_\lambda$. By transferring $f_{\lambda,j}$ across the tensor product one obtains the following decomposition of $\rmi H_{\rm N}$
\begin{equation}
\label{h_lambda}
\rmi H_{\rm N}=\bigoplus_{\lambda \in \hat G_0^{\rm r}} 1 \otimes_{F_\lambda} h_\lambda,
\end{equation}
where $h_\lambda: E_\lambda \to E_\lambda$ is given by
$$
h_\lambda = \sum_j f_{\lambda,j} \zeta_{\lambda,j} .
$$
The comparison of (\ref{g_action}) and (\ref{h_lambda}) reveals that the action of $g \in G_0$ and that of $h_\lambda$ are completely decoupled. Therefore, no symmetry conditions are imposed on $h_\lambda$, except that of the commutation with the scalars of $F_\lambda$. Recall, however, that we have not yet taken into account the possible antiunitary symmetries of the system. As we shall see in section \ref{sec:antiunitary}, they give rise to additional
constraints on $h_\lambda$.
\section{Antiunitary and chiral symmetries}\label{sec:antiunitary}
In this section we first characterize the ${\cal Q}\mbox{-compatible}$ antiunitary operators in $\wedge V$. Then we define their (projective) representation in the Nambu space. Finally, we describe the action of antiunitary operators on the isotypic components of (\ref{isotypic}).
\par
Let us first consider the important case of antiunitary operator $C_{\rm F}$ in the Fock space. It is defined as a complex Hodge dual:
\begin{equation}
\label{C_F}
\Omega \cdot \langle C_{\rm F} x, y \rangle  = x \wedge y,
\end{equation}
where $x, y \in \wedge V$ and $\Omega$ is a fixed fully occupied state of unit norm.
Let $\sigma$ be a permutation of the set $\{1,\dots,N\}$. The action (\ref{C_F}) of $C_{\rm F}$ on the canonical basis of the Fock space can be written explicitly
\begin{equation}
\label{action_cf}
C_{\rm F}\left( a^\dag_{\sigma(1)}\dots a^\dag_{\sigma(n)}\left|0\right\rangle\right)=\sgn(\sigma) \left( a^\dag_{\sigma(n+1)}\dots a^\dag_{\sigma(N)}\left|0\right\rangle\right),
\end{equation}
where $0\le n\le N$. For an arbitrary $n\mbox{-particle}$ state $|\psi\rangle$ the following relations hold:
\begin{eqnarray}
\label{commutation_cf}
\left(C_{\rm F} a^\dag_{k} C_{\rm F}^{-1} \right) \left|\psi\right\rangle =
(-1)^{n}a_{k} \left|\psi\right\rangle\\\ \nonumber
\left(C_{\rm F} a_{k}C_{\rm F}^{-1} \right)\left|\psi\right\rangle=
(-1)^{n-1}a^\dag_{k} \left|\psi\right\rangle,
\end{eqnarray}
thus $C_{\rm F}$ can be interpreted as a charge conjugation operator in the Fock space. It is convenient to introduce a twisted version of $C_{\rm F}$:
\begin{equation}
\label{CFU}
\tilde C_{\rm F} = C_{\rm F} U,
\end{equation}
where the unitary operator $U$ is defined by
\begin{equation}
\label{U}
U=\exp\left(\frac{\rmi\pi}{2}\sum_k \left(a^\dag_k a_k-\frac{1}{2}\right)\right).
\end{equation}
One can check that for any $X_{\rm F} \in {\cal Q}$
\begin{equation}
\label{cf_anticomm}
\tilde C_{\rm F} X_{\rm F} \tilde C_{\rm F}^{-1} = - X_{\rm F}.
\end{equation}
Therefore, $\tilde C_{\rm F}$ is ${\cal Q}\mbox{-compatible}$.
\par
One might be tempted to extend the linear representation of $G_{{\rm F}0}$ in the Nambu space defined in section \ref{sec:nambu} to $G_{{\rm F}1}=G_{\rm F}\setminus G_{{\rm F}0}$. However, this is not possible, as can be seen from the example of $\tilde C_{\rm F}$. Indeed, since $\tilde C_{\rm F}$ is antiunitary, equation (\ref{cf_anticomm}) implies that it commutes with $\rmi X_{\rm F}$ for all $X_{\rm F} \in {\cal Q}$. That is, the corresponding Nambu operator $\tilde C_{\rm N}$ would commute with all generators of $\mathrm{SO}(W_\R)$. Therefore, $\tilde C_{\rm N}$ equals plus or minus identity in $W_\R$, yielding in either case $\tilde C_{\rm N}^2=1$. On the other hand, formula (\ref{action_cf}) shows that $\tilde C_{\rm F}$ squares to
\begin{equation}
\label{cf2}
\tilde C_{\rm F}^2=\begin{cases}
1 & \text{for $N$ odd,}\\
(-1)^{N/2} U^2 & \text{for $N$ even.}
\end{cases}
\end{equation}
Since $U^2$ is represented in $W_\R$ by $-1$, equation (\ref{cf2}) would imply $\tilde C_{\rm N}^2=-1$ for $N$ even, in contradiction with the above result. If $N$ is odd, a similar reasoning applies to the operator $\tilde C_{\rm F}(a_{e_0}+a^\dag_{e_0})$. Therefore, one cannot extend the linear representation of $G_{{\rm F}0}$ in $W_\R$ to antiunitary operators. It is still possible to define {\em a projective representation} of antiunitary symmetries in the Nambu space by setting $\tilde C_{\rm N}=1$ and using the fact that any antiunitary operator $A_{\rm F}$ is a product of $\tilde C_{\rm F}$ and a unitary operator. We shall use, however, a different construction.
\par
$U^2$ leaves invariant the subspaces $\wedge^n V$, where it acts by multiplication by $\rmi^{2n-N}$. Therefore, $U^2$ commutes with any parity-conserving operator in $\wedge V$, including the hamiltonian. We shall thus assume that $U^2$ always belongs to the subgroup $G_{{\rm F}0}$ of unitary symmetries. Let us define the map $\chi: G_{{\rm F}1} \to \mathrm{U}(\wedge V)$ by
\begin{equation}
\label{chi}
\chi: A_{\rm F}\mapsto \tilde C_{\rm F} A_{\rm F}.
\end{equation}
Clearly, $\chi(A_{\rm F})$ is a ${\cal Q}\mbox{-compatible}$ unitary operator in the Fock space, hence $\chi(A_{\rm F}) \in \mathrm{Pin}^\C(2N)$. Let us define a subset $\tilde G_{\rm F} \subset \mathrm{Pin}^\C(2N)$ in the following way:
\begin{equation}
\label{tilde_G}
\tilde G_{\rm F}=G_{{\rm F}0} \cup \chi(G_{{\rm F}1}).
\end{equation}
We claim that $\tilde G_{\rm F}$ is a subgroup of $\mathrm{Pin}^\C(2N)$. This stems from the fact that $\tilde C_{\rm F}$ commutes with all parity-conserving unitary operators $\exp\left(\rmi(X_{\rm F}+c)\right)$ and that for the parity-changing operator $a_{e_0}+a^\dag_{e_0}$ one has
$$
\tilde C_{\rm F}\left(a_{e_0}+a^\dag_{e_0}\right)\tilde C_{\rm F}^{-1}=\rmi^{N+1} U^2 \left(a_{e_0}+a^\dag_{e_0}\right).
$$
On the other hand, formula (\ref{cf2}) shows that $\tilde C_{\rm F}^2 \in G_{{\rm F}0}$. Therefore, the subset $\tilde G_{\rm F}\subset \mathrm{Pin}^\C(2N)$ is closed with respect to multiplication and inversion in $\mathrm{Pin}^\C(2N)$.
\par
It should be emphasized that group $\tilde G_{\rm F}$ defined in (\ref{tilde_G}) is {\em not} the symmetry group of the system. Indeed, for any antiunitary symmetry $A_{\rm F}$, the corresponding unitary operator $\chi(A_{\rm F})$ anticommutes with the hamiltonian:
\begin{equation}
\label{chiral}
\chi(A_{\rm F}) H \chi(A_{\rm F})^{-1} = -H.
\end{equation}
The unitary operators satisfying (\ref{chiral}) are often referred to as ``chiral symmetries''. The above considerations show that  antiunitary symmetry constraints on $H$ can be equivalently replaced by unitary chiral symmetries of the form (\ref{chiral}). Since group $\tilde G_{\rm F}$ consists of unitary operators only, it admits a Nambu space representation. Let us define $G \subset \mathrm{O}(W_\R)$ as a factor of $\tilde G_{\rm F}$ by the kernel of its action on $W_\R$:
\begin{equation}
\label{G}
G=\tilde G_{\rm F} / \mathrm{U}(1).
\end{equation}
Throughout the rest of the paper we will work exclusively with group $G$ and its representation in $W_\R$.
\par
The group $G$ defined by (\ref{G}) contains $G_0$ as a normal subgroup of index 2. Let $G_1$ stand for the coset of $G_0$ corresponding to the Nambu space representation of $\chi(G_{{\rm F}1})$. We are now ready to describe the action of the elements of $G_1$ on isotypic components in (\ref{isotypic}):
\begin{corollary}
\label{cor2}
The action of an element $s \in G_1$ on $W_\R$ respects decomposition~(\ref{isotypic}):
\begin{equation}
\label{s1}
s({\cal W}_{\lambda})={\cal W}_{\lambda'}.
\end{equation}
If $\lambda \neq \lambda'$, the blocks ${\cal W}_{\lambda}$ and ${\cal W}_{\lambda'}$ are transposed:
\begin{equation}
\label{s2}
s({\cal W}_{\lambda'})={\cal W}_{\lambda}
\end{equation}
and the algebras of intertwining operators for the representations ${\cal R}_{\lambda}$ and ${\cal R}_{\lambda'}$ are isomorphic:
$$
F_{\lambda}=F_{\lambda'}.$$
The action of $s$ on each isotypic component (\ref{lambda_block}) admits a tensor product factorization
\begin{equation}
\label{pure_tensor}
s\Bigr|_{{\cal W}_\lambda}=\xi_{s, \lambda} \otimes_{F_{\lambda}} \phi_{\lambda},
\end{equation}
where $\phi_{\lambda}: E_{\lambda} \to E_{\lambda'}$ does not depend on $s$. Both $\xi_{s, \lambda}$ and $\phi_{\lambda}$ are $F_\lambda\mbox{-linear}$ except for the case $F_\lambda=\C$, where both $\xi_{s, \lambda}$ and $\phi_{\lambda}$ might be $\C\mbox{-antilinear}$. In the case $\lambda = \lambda'$, $\phi_{\lambda}$ squares to plus or minus identity:
$$
\phi_{\lambda}^2=\pm 1_{E_\lambda},
$$
except when $F_\lambda=\C$ and both $\xi_{s, \lambda}$ and $\phi_{\lambda}$ are $\C\mbox{-linear}$, in which case $\phi_{\lambda}^2=+1_{E_\lambda}$.
\end{corollary}
The proof of the above is given in the Appendix.
\par
So far we treated all antiunitary symmetries on an equal footing. However, there are situations, where one of the elements of $G_1$ belongs to the center of $G$. Following Dyson \cite{dyson} we shall interpret the action of this element as a time reversal operator $T$ (see also \cite{sachs}). Since $T$ commutes with any $g \in G_0$, the mapping $\xi_{T, \lambda}$ acts on ${\cal R}_\lambda$ by multiplication by some $f \in F_\lambda$. On the other hand, since $\xi_{T, \lambda}$ is $F_\lambda\mbox{-linear}$ in the case $F_\lambda=\HH$, $f$ commutes with all quaternions and must be a real number. The same is obviously true for $F_\lambda=\R$, hence $\xi_{T, \lambda}^2=+1$ for both cases $F_\lambda=\R$ and $F_\lambda=\HH$. Therefore, in these two cases the sign of $\phi_\lambda^2$ coincides with that of $T^2$ in the Nambu space. In the case $F_\lambda=\C$, however, both cases $\xi_{T, \lambda}^2=+1$ and $\xi_{T, \lambda}^2=-1$ are possible(see \cite{hhz} for detailed explanation of the change of sign of $T^2$ between the Nambu space representation and the multiplicity space).
\section{Spectrum gaps and Clifford modules}\label{sec:gaps}
We will now apply the techniques developed in the previous section to the problem of classification of gapped quadratic hamiltonians. Let us first make use of the faithfulness of the Nambu representation of ${\cal Q}$ and reformulate the problem entirely in $W_\R$. In particular, instead of deriving $G_0$ and $G$ from the symmetries of the Fock space by diagram (\ref{GF0_G0}) and formula (\ref{G}), we shall interpret these groups as primary symmetries of the system:
\begin{quote}
Let $W_\R$ be a real vector space with the Euclidean norm and the orthogonal action of a compact group $G$. The group $G$ contains a subgroup $G_0$. Two cases are considered:
\begin{itemize}
\item{Either $G_0$ coincides with $G$}.
\item{Or $[G:G_0]=2$, that is $G_0$ is a proper subgroup of index 2 in $G$. In this case $G_1$ stands for the complement of $G_0$ in $G$.}
\end{itemize}
Let ${\cal S}$ be the space of real antisymmetric operators $\rmi H_{\rm N}$ in $W_\R$ with the following properties:
\begin{itemize}
\item{$\rmi H_{\rm N}$ commutes with the action of $G_0$ and anticommutes with the action of the elements of $G_1$ (if $G_1 \neq \emptyset$).}
\item{$0 \notin \mathrm{Spec}\left(\rmi H_{\rm N}\right)$, in the other words, the spectrum of $\rmi H_{\rm N}$ has a gap at zero energy.}
\end{itemize}
What is the homotopy type of ${\cal S}$?
\end{quote}
Note that no distinction is now made between antiunitary and chiral symmetries, since the elements of $G_1$ represent both types of operators indiscriminately.
\par
Let us first ``flatten'' the spectrum of $\rmi H_{\rm N}$. The function $y(x, t): \C\times[0,1] \to \C$ defined by the formula
\begin{equation}
\label{flatten}
y(x,t)=(1-t)x+\rmi t\,\sgn\left(\mathrm{Im}(x)\right)
\end{equation}
is continuous on $\mathrm{Spec}(\rmi H_{\rm N}) \times [0,1]$ and odd with respect to $x$ for every $t \in [0,1]$. Since $\rmi H_{\rm N}$ is a normal operator, one can define an operator-valued function ${\cal O}(t)=y(\rmi H_{\rm N}, t)$, such that ${\cal O}(0)=\rmi H_{\rm N}$ and ${\cal O}(t)$ satisfies the above constraints for every $t \in [0,1]$. Let us define the `flattened'' hamiltonian $\tilde H_{\rm N}$ by $\rmi \tilde H_{\rm N}={\cal O}(1)$ and denote by $\tilde {\cal S}$ the space of operators $\rmi \tilde H_{\rm N}$. It is clear from the above that $\tilde {\cal S}$ is a deformation retract of $\cal S$.
\par
$\rmi \tilde H_{\rm N}$ admits a decomposition similar to (\ref{h_lambda}):
\begin{equation}
\label{flat_h}
\rmi \tilde H_{\rm N}=\bigoplus_{\lambda \in \hat G_0^{\rm r}} 1 \otimes_{F_\lambda} \tilde h_\lambda.
\end{equation}
Since the spectrum of $\rmi \tilde H_{\rm N}$ consists of the two points $\{-\rmi, \rmi\}$, one has $\tilde h_\lambda^2=-1$ for every $\lambda \in \hat G_0^{\rm r}$. Our strategy will consist in splitting the problem into parts by considering each $\tilde h_\lambda$ in (\ref{flat_h}) independently. However, some isotypic components ${\cal W}_\lambda$ and ${\cal W}_{\lambda'}$ can be transposed by the action of $G_1$. In this case $\tilde h_\lambda$ and $\tilde h_{\lambda'}$ are not independent. Since $\rmi \tilde H_{\rm N}$ anticommutes with the elements of $G_1$, one has indeed:
\begin{equation}
\label{h_clone}
\tilde h_{\lambda'}=-\phi_\lambda \tilde h_\lambda \phi_\lambda^{-1},
\end{equation}
where $\phi_\lambda$ is defined in (\ref{pure_tensor}). Let us consider the action of $G_1$ on $\hat G_0^{\rm r}$. The orbits of this action are of length 1 or 2. Let us choose a subset $\hat G_0^{\rm ra} \subset \hat G_0^{\rm r}$ in such a way that each orbit contains exactly one point of $\hat G_0^{\rm ra}$. The operators $\tilde h_\lambda$ for all $\lambda \in \hat G_0^{\rm ra}$ are independent, and it follows from (\ref{h_clone}) that all terms in the decomposition (\ref{flat_h}) are completely determined by them. Therefore, $\tilde {\cal S}$ has the structure of a direct product:
\begin{equation}
\label{S_product}
\tilde {\cal S}=\prod_{\lambda \in \hat G_0^{\rm ra}} \tilde {\cal S}_\lambda,
\end{equation}
where $\tilde {\cal S}_\lambda$ stands for the space of operators $\tilde h_\lambda$.
\par
Let us now describe the structure of $\tilde {\cal S}_\lambda$. It is determined by the constraints imposed on the operators $\tilde h_\lambda: E_\lambda \to E_\lambda$. Since the constraint of $G_0\mbox{-equivariance}$ is taken into account, there remain two conditions: the commutation with scalars
\begin{equation}
\label{h_f}
\tilde h_\lambda f = f \tilde h_\lambda
\end{equation}
for any $f \in F_\lambda$, and the anticommutation with the elements of $G_1$. The latter applies only when $G_1$ is not empty and when its elements map the isotypic component ${\cal W}_\lambda$ to itself. This condition then reads
\begin{equation}
\label{h_phi}
\tilde h_\lambda \phi_\lambda = - \phi_\lambda \tilde h_\lambda.
\end{equation}
\par
Equations (\ref{h_f}) end (\ref{h_phi}) together with the rules of the commutation of $\phi_\lambda$ with $F_\lambda$ determine the joint action of $F_\lambda$, $\phi_\lambda$ and $\tilde h_\lambda$ on $E_\lambda$ completely.
Ten cases are possible: three correspond to the situation when $G=G_0$, the other seven cases are described in  Corollary 2. It turns out that in all cases the joint action of $F_\lambda$, $\phi_\lambda$ and $\tilde h_\lambda$ extends to the action of one of the ten Clifford algebras. Recall that the real Clifford algebra associated with a non-degenerate quadratic form is characterized up to isomorphism by its signature. For the quadratic form with $p$ negative and $q$ positive eigenvalues, the corresponding algebra $C^{p,q}$ is described explicitly by generators $\{\epsilon_1,\dots,\epsilon_{p+q}\}$ and relations
\begin{eqnarray}
\epsilon_k^2=
\begin{cases}
-1& \text{for $1\le k \le p$} \\
$+1$& \text{for $p+1 \le k \le p+q$}
\end{cases}
\nonumber \\
\epsilon_k \epsilon_m = -\epsilon_m \epsilon_k\quad\mbox{ if }k \neq m. \nonumber
\end{eqnarray}
The complex Clifford algebras $C^n_\C$ are defined in a similar way, except that in the complex case the algebra is characterized by a single index $n$ since all generators can be chosen positive. More details on the Clifford algebras can be found in \cite{heat_kernel, karoubi, spin_geometry}.
\par
The explicit action of the Clifford algebra generators on $E_\lambda$ is given in Tables \ref{tab:real} and \ref{tab:complex}. We follow here the approach of \cite{kitaev} and interpret the action of the Clifford algebra generated by $F_\lambda$ and $\phi_\lambda$ alone as a constraining framework for the hamiltonian. This algebra is denoted $C^{p,q}$ in Table \ref{tab:real} and $C^n_\C$ in Table \ref{tab:complex}. The action of $\tilde h_\lambda$ extends the above algebra in one of the following ways:
\begin{equation}
\label{extensions}
\begin{aligned}
C^{p,q}&\to C^{p+1, q};\\
C^{p,q}&\to C^{p, q+1};\\
C^n_\C&\to C^{n+1}_\C .
\end{aligned}
\end{equation}
Conditions (\ref{extensions}) allow one to identify the Clifford algebras $C^{p,q}$ and $C^n_\C$ unambiguously despite the occasional unnatural isomorphisms $C^{1,0}=C^0_\C$ and $C^{1,1}=C^{0,2}$.
\par
\begin{table}
\caption{\label{tab:real}Extensions of the real Clifford algebra $C^{p,q}$ by $\tilde h_\lambda$. The real $K\mbox{-theory}$ index $n$ is given by formula (\ref{n_real}), $\phi_\lambda$ is defined in (\ref{pure_tensor}). The symbols $i$, $j$ and $k$ stand for the standard imaginary units in $F_\lambda=\C$ or $F_\lambda=\HH$; in the last case $i^2=j^2=k^2=-1$ and $ij=-ji=k$. The element $\tilde h_\lambda$ is defined by (\ref{flat_h}). The meaning of the Cartan labels is explained in section \ref{sec:gaps}.}
{\begin{tabular}{@{}*{7}{l}}
\hline\\
&&&Generators&Extending&Clifford algebra&Cartan\cr
$n$&$F_\lambda$&$\phi_\lambda^2$&of $C^{p,q}$&generator&extension&label\cr
\hline\\
0&$\C$&$+1$&$\{i,\,\phi_\lambda\}$&$\phi_\lambda \tilde h_\lambda$&$C^{1,1} \to C^{1,2}$&AI\cr
1&$\R$&$+1$&$\{\phi_\lambda\}$&$\tilde h_\lambda$&$C^{0,1} \to C^{1,1}$&BDI\cr
2&$\R$&---&$\emptyset$&$\tilde h_\lambda$&$C^{0,0} \to C^{1,0}$&D\cr
3&$\R$&$-1$&$\{\phi_\lambda\}$&$\tilde h_\lambda$&$C^{1,0} \to C^{2,0}$&DIII\cr
4&$\C$&$-1$&$\{i,\,\phi_\lambda\}$&$\phi_\lambda \tilde h_\lambda$&$C^{2,0} \to C^{3,0}$&AII\cr
5&$\HH$&$+1$&$\{i,\,j,\,k\phi_\lambda\}$&$k\tilde h_\lambda$&$C^{3,0} \to C^{3,1}$&CII\cr
6&$\HH$&---&$\{i,\,j\}$&$k\tilde h_\lambda$&$C^{2,0} \to C^{2,1}$&C\cr
7&$\HH$&$-1$&$\{i,\,j,\,k\phi_\lambda\}$&$k\tilde h_\lambda$&$C^{2,1} \to C^{2,2}$&CI\cr
\hline
\end{tabular}}
\end{table}
\begin{table}
\caption{\label{tab:complex}Extensions of the complex Clifford algebra $C_\C^n$ by $\tilde h_\lambda$. See Table \ref{tab:real} for more details. Note that in the case $n=1$, $\phi_\lambda^2$ can be made equal to any element of $F_\lambda$ of unit norm.}
{\begin{tabular}{@{}*{7}{l}}
\hline\\
&&&Generators&Extending&Clifford algebra&Cartan\cr
$n$&$F_\lambda$&$\phi_\lambda^2$&of $C_\C^n$&generator&extension&label\cr
\hline
0&$\C$&---&$\emptyset$&$\tilde h_\lambda$&$C_\C^0 \to C_\C^1$&A\cr
1&$\C$&$+1$&$\{\phi_\lambda\}$&$\tilde h_\lambda$&$C_\C^1 \to C_\C^2$&AIII\cr
\hline
\end{tabular}}
\end{table}
\par
We can now characterize $\tilde {\cal S}_\lambda$ for all  cases listed in Tables \ref{tab:real} and \ref{tab:complex}. To distinguish between them we shall use Cartan labels; the exact meaning of these symbols will be explained below. Let $\R(d)$, $\C(d)$ and $\HH(d)$ stand respectively for the algebras of real, complex and quaternionic $d \times d$ matrices. It is well known (see e.g. \cite{karoubi}) that any simple real or complex Clifford algebra is isomorphic to $\R(d)$, $\C(d)$ or $\HH(d)$. When the Clifford algebra is not simple (this is the case for $C^{p, q}$ with $p-q=3 \bmod 4$ and for $C^n_\C$ when $n$ is odd), it is isomorphic to the direct sum of two identical algebras of one of the above types. Eventually, there are three possibilities:
\begin{itemize}
\item{The extension by $\tilde h_\lambda$ transforms a non-simple algebra to a simple one. This is the case for classes BDI, AIII and CII. Let us consider the symmetry class BDI in details. The algebra homomorphism (\ref{extensions}) reads in this case $\R\oplus\R\to\R(2)$ and corresponds to the inclusion of the algebra of real diagonal $2\times 2$ matrices into $\R(2)$. As an $\R(2)\mbox{-module}$, $E_\lambda$ can be thought of as $\R^k\otimes_\R \R^2$, while its structure as an $\R\oplus\R\mbox{-module}$ looks like $\R^k\oplus\R^k$. Therefore, the action of $\tilde h_\lambda$ is a norm-preserving mapping of $\R^k$ to $\R^k$, hence $\tilde {\cal S}_\lambda=\mathrm{O}(k)$. The similar reasoning applies to classes AIII and CII, yielding
\begin{equation}
\label{sum_to_simple}
\tilde {\cal S}_\lambda=\begin{cases}
\mathrm{O}(k) & \text{for BDI ($E_\lambda=\R^{2k}$)}\\
\mathrm{U}(k) & \text{for AIII ($E_\lambda=\C^{2k}$)}\\
\mathrm{Sp}(k) & \text{for CII ($E_\lambda=\HH^{2k}$)}
\end{cases}
\end{equation}
}
\item{The extended algebra is a sum of two simple ones. This occurs for classes AI, A and AII. In the case AI, the homomorphism (\ref{extensions}) looks like $\R(2) \to \R(2)\oplus\R(2)$ and corresponds to the mapping of $2 \times 2$ real matrices to block-diagonal $4 \times 4$ matrices with two identical $2 \times 2$ blocks. As a module over $\R(2)$, the space $E_\lambda$ has the structure $\R^k\otimes_\R \R^2$ (note that as an $F_\lambda\mbox{-module}$ $E_\lambda=\C^k$). The action of $\tilde h_\lambda$ splits $\R^k$ into two blocks $\R^m$ and $\R^{k-m}$. This splitting is first parameterized by the dimensions of the blocks and second by the corresponding Grassmannian manifolds. The cases A and AII are similar, except that the Grassmannians are the complex and the quaternionic ones respectively:
\begin{equation}
\label{simple_to_sum}
\tilde {\cal S}_\lambda=\begin{cases}
\bigcup_{m=0}^k \mathrm{O}(k)/\left(\mathrm{O}(m)\times \mathrm{O}(k-m)\right) & \text{for AI ($E_\lambda=\C^k$)}\\
\bigcup_{m=0}^k \mathrm{U}(k)/\left(\mathrm{U}(m)\times \mathrm{U}(k-m)\right) & \text{for A ($E_\lambda=\C^k$)}\\
\bigcup_{m=0}^k \mathrm{Sp}(k)/\left(\mathrm{Sp}(m)\times \mathrm{Sp}(k-m)\right) & \text{for AII ($E_\lambda=\C^{2k}$)}
\end{cases}
\end{equation}
}
\item{Both Clifford algebras are simple. This is the case for classes D, C, CI and DIII. In the case D, for instance, the homomorphism (\ref{extensions}) corresponds to the inclusion $\R \to \C$. Thus, the action of $\tilde h_\lambda$ corresponds to assigning a complex structure to the real vector space $E_\lambda=\R^{2k}$. Since the complex structure is invariant under unitary transformations, one has $\tilde {\cal S}_\lambda=\mathrm{O}(2k)/\mathrm{U}(k)$. Cases C, CI and DIII can be treated in a similar way. One obtains the following formulas:
\begin{equation}
\label{simple_to_simple}
\tilde {\cal S}_\lambda=\begin{cases}
\mathrm{O}(2k)/\mathrm{U}(k) & \text{for D ($E_\lambda=\R^{2k}$)}\\
\mathrm{Sp}(k)/\mathrm{U}(k) & \text{for C ($E_\lambda=\HH^k$)}\\
\mathrm{U}(k)/\mathrm{O}(k) & \text{for CI ($E_\lambda=\HH^k$)}\\
\mathrm{U}(2k)/\mathrm{Sp}(k) & \text{for DIII ($E_\lambda=\R^{4k}$)}
\end{cases}
\end{equation}
}
\end{itemize}
\par
When the dimension $k$ of $C^{p, q}\mbox{-module}$ $E_\lambda$ is small, $\tilde
{\cal S}_\lambda$ may exhibit exceptional homotopy properties, which disappear
with increasing $k$. This phenomenon is known as the stabilization of the
homotopy type \cite{at_handbook}. As an example one may consider the symmetry
class AI with $k=2$. In this case, according to (\ref{simple_to_sum}), $\tilde
{\cal S}_\lambda$ is the union of two points and a circle. Therefore, the
homotopically nontrivial mappings $S^1 \to \tilde {\cal S}_\lambda$ are
characterized in this case by an integer from $\pi_1(S^1)=\Z$, while for $k>3$
the adiabatic cycles in the quantum dot of the class AI are labeled by $\Z_2$~\cite{roy}. Increasing the dimension $k$ corresponds to augmenting the system by a set of local modes \cite{kitaev}. The example above shows that such augmentation can modify the situation.
\par
The appropriate mathematical tools for the study of the stable homotopy properties of $\tilde {\cal S}_\lambda$ are provided by $K\mbox{-theory}$. There are two ways to construct the $K\mbox{-groups}$ for the problem under consideration. One is purely algebraic and based on the consideration of the equivalence classes of formal differences between $C^{p, q}\mbox{-modules}$ \cite{kitaev}. The other is more geometrical and physically intuitive. This approach consists in working with the inductive limits of $\tilde {\cal S}_\lambda$ with respect to the increasing dimension of $E_\lambda$. These limits are known as the classifying spaces of $K\mbox{-theory}$, and depend only on Morita equivalence classes of $C^{p,q}$ \cite{karoubi}. In the case of the real Clifford algebras the classifying spaces are denoted by $R_n$, where the index $n$ is given by the formulas
\begin{equation}
\label{n_real}
\begin{aligned}
n=q-p \bmod 8\qquad& \mbox{for }  C^{p,q}\to C^{p,q+1} \cr
n=p-q+2 \bmod 8\qquad& \mbox{for }  C^{p,q}\to C^{p+1,q}
\end{aligned}
\end{equation}
for the positive and negative extensions respectively. In the case of the extension of complex Clifford algebras $C^m_\C \to C^{m+1}_\C$ the corresponding classifying space depends only on $n = m \bmod 2$ and is denoted by $C_n$.
\par
Let us now explain the meaning of the Cartan labels in Tables \ref{tab:real} and \ref{tab:complex}. It has been observed in \cite{az} that there is a correspondence between the symmetry classes of the fermionic hamiltonians and some infinite series of compact Lie groups and symmetric spaces. According to \cite{hhz}, the space of the time evolution operators for the system of a given symmetry class always belongs to one of these series. This fact can be easily understood within the framework of multiplicity spaces (note that for these matters the spectrum of the hamiltonian does not need to have a gap).
\par
Let us fix a $G_0\mbox{-invariant}$ quadratic form $r \mapsto |r|^2$ on ${\cal R}_\lambda$. Since ${\cal R}_\lambda$ is irreducible, any other $G_0\mbox{-invariant}$ form on ${\cal R}_\lambda$ differs from $|r|^2$ only by a factor. Therefore, for any $e \in E_\lambda$ the function $r \mapsto |e(r)|^2$ is proportional to $|r|^2$ and the expression
\begin{equation}
\label{form_e}
|e|^2=\frac{|e(r)|^2}{|r|^2}
\end{equation}
defines a real valued quadratic form on $E_\lambda$. The form (\ref{form_e}) is clearly invariant with respect to the multiplication by unitary elements of $F_\lambda$ and to the action of the evolution operator $u_t$ defined by
$$
u_t=\exp(h_\lambda t).
$$
The polarization identity transforms (\ref{form_e}) to an inner product on $E_\lambda$ (cf \cite{karow} for the details on the polarization identity in quaternionic vector spaces). The operator $u_t$ is $F_\lambda\mbox{-linear}$ and conserves the inner product. Therefore, it belongs to one of the compact simple Lie groups ${\cal G}=\mathrm{SO}(E_\lambda)$, $\mathrm{U}(E_\lambda)$ or $\mathrm{Sp}(E_\lambda)$. In the Cartan classification \cite{cartan1} they correspond to the infinite series ${\rm D}_n$, ${\rm A}_n$ and ${\rm C}_n$, which explains the choice of the labels for the corresponding symmetry classes. Antiunitary and chiral symmetries impose additional constraints on the space of operators $u_t$. Specifically, $u_t$ must satisfy the condition
$$
\phi_\lambda u_t \phi_\lambda^{-1} = u_t^{-1}.
$$
The above equation defines the Cartan embedding \cite{cartan_embedding} of the symmetric space ${\cal G}/{\cal H}$ into ${\cal G}$, where ${\cal H}$ is a subgroup of $\phi_\lambda\mbox{-invariant}$ elements of ${\cal G}$. There are seven infinite series of compact symmetric spaces \cite{cartan2}, and all of them appear in the classification of fermion hamiltonians \cite{hhz}.
\par
Let us illustrate the above construction by the case of classes CI and CII in Table \ref{tab:real}. For both classes $F_\lambda=\HH$. Let us consider first the case $\phi_\lambda^2=-1$. In this case $\phi_\lambda$ defines a complex structure in the quaternionic vector space $E_\lambda$ and the $\phi_\lambda\mbox{-invariant}$ operators form a complex unitary group. Thus the symmetric space ${\cal G}/{\cal H}$ is isomorphic to $\mathrm{Sp}(n)/\mathrm{U}(n)$ and belongs to type CI \cite{helgason}. On the other hand, in the case $\phi_\lambda^2=+1$, the space $E_\lambda$ can be decomposed into a sum of two $\phi_\lambda\mbox{-invariant}$ subspaces with eigenvalues $+1$ and $-1$. ${\cal H}$ is then a product of two quaternionic unitary groups, and ${\cal G}/{\cal H}$ is of type CII.
\section{Examples}\label{sec:examples}
\subsection{Normal systems}
Let us consider the normal (non superconducting) systems, corresponding to $B=0$ in equation (\ref{traceless_quadratic}). In this case the Nambu space representation is redundant, since the conservation of the number of particles makes it possible to describe the system in the single particle space $V$. However, the general results of section \ref{sec:gaps} remain applicable.
\par
Let us consider the operator $J:W_\R \to W_\R$:
\begin{equation}
\label{J}
J:\,e+\langle e, \cdot\rangle \mapsto \rmi e -\rmi\langle e, \cdot\rangle.
\end{equation}
Clearly, $J^2=-1$. Therefore, $J$ defines a complex structure on the space $W_\R$  inherited from that of $V$. In the case of normal systems, $J$ commutes with $\rmi H_{\rm N}$, therefore for any $\varphi \in \R$ the operator
$$
u_\varphi=\exp(J\varphi)
$$
belongs to $G_0$. However, the action of $u_\varphi$ on the space $V$ consists in the multiplication by a global phase factor $\rme^{\rmi\varphi}$. Traditionally, multiplication by scalars is not included in the group of unitary symmetries of a single particle system. We will denote the latter group by $G_0'$ to distinguish it from the full group $G_0$ of unitary symmetries in the Nambu space\footnote{In Dyson's notations \cite{dyson} the group algebras of $G_0'$ and $G_0$ are called $A$ and $B$ correspondingly.}. $G_0$ is therefore generated by $G_0'$ and $u_\varphi$. Consider now (\ref{isotypic}) the isotypic decomposition of $W_\R$. Since $J$ commutes with the actions of both $G_0'$ and $u_\varphi$, it corresponds to an intertwining operator of ${\cal R}_\lambda$ for all isotypic components of (\ref{isotypic}). Therefore, $F_\lambda$ contains at least one imaginary unit. On the other hand, $J$ also belongs to $\R G_0$, hence it commutes with all elements of $F_\lambda$. This is only possible if $F_\lambda=\C$, leaving us with the following four symmetry classes: A, AI, AII and AIII.
\par
Let us now consider antiunitary or chiral symmetries. First of all, if none of them is present, the system belongs to class A. To distinguish between the remaining cases, let us consider the conjugation of $J$ by an element $s \in G_1$. Since the operator $\xi_{s, \lambda}$ defined by (\ref{pure_tensor}) is $F_\lambda\mbox{-antilinear}$ for cases AI and AII and linear for case AIII, one has
\begin{equation}
\label{sJs}
sJs^{-1}=\begin{cases}
-J & \text{for AI or AII,}
\\ J & \text{for AIII.}
\end{cases}
\end{equation}
So far no distinction has been made between antiunitary and normal symmetries in the Nambu space representation. However, since the mapping (\ref{chi}) involves the twisted charge conjugation operator $\tilde C_{\rm F}$, it does not respect the conservation of the number of particles. Therefore, in the case of normal systems, one can discriminate two types of symmetry. Indeed, since $J$ is inherited from the complex structure of $V$, it follows from equation (\ref{sJs}) that the elements of $G_1$ correspond to chiral symmetries in the case of class AIII and to antiunitary operators in the cases AI and AII.
\par
Let us now analyze the case when a normal system possesses the time reversal symmetry $T$. The corresponding $\R\mbox{-linear}$ operator $\xi_{T, \lambda}:{\cal R}_\lambda \to {\cal R}_\lambda$ commutes with all elements of $G_0'$. Since $\xi_{T, \lambda}$ and $J$ anticommute with each other, they generate an algebra isomorphic to $\R(2)$ when $\xi_{T, \lambda}^2=1$ and to $\HH$ when $\xi_{T, \lambda}^2=-1$. Therefore the following situations are possible:
\begin{itemize}
\item{$G_0'$ acts on ${\cal R}_\lambda$ by a representation of real type. Then $T^2=1$ for class AI and $T^2=-1$ for class AII.}
\item{$G_0'$ acts on ${\cal R}_\lambda$ by a representation of quaternionic type. Then $T^2=1$ for class AII and $T^2=-1$ for class AI.}
\end{itemize}
In \cite{dyson} the corresponding cases are referred to as {\it RR}, {\it QR}, {\it QQ} and {\it RQ}.
\subsection{Dirac hamiltonians}
Although the relevant symmetry group for Dirac hamiltonians includes arbitrary
translation and thus is not compact, the above classification scheme can be
straightforward applied \cite{kitaev}\ to the mass term of the Dirac equation. Let us consider a generalized Dirac hamiltonian in the Majorana basis:
\begin{equation}
\label{Dirac}
\hat H=-\rmi\int  \hat c^i_\mathbf{x}\left(M_{ij} + \gamma_{ij}^\alpha \partial_\alpha\right)\hat c^j_\mathbf{x} \rmd^d\mathbf{x},
\end{equation}
where $\hat c^i_\mathbf{x}$ are $2N\mbox{-component}$ Majorana fermion fields:
\begin{equation}
\begin{aligned}
\hat c^{i\dag}_\mathbf{x}&=\hat c^i_\mathbf{x}\\
\left\{\hat c^i_\mathbf{x}, \hat c^j_\mathbf{y}\right\}&=2\delta^{ij}\delta(\mathbf{x}-\mathbf{y}).
\end{aligned}
\end{equation}
The real symmetric matrices $\gamma_{ij}^\alpha$ define a representation of the Clifford algebra $C^{0,d}$ in $\R^{2N}$:
\begin{equation}
\label{gamma}
\gamma_{ij}^\alpha \gamma_{jk}^\beta + \gamma_{ij}^\beta \gamma_{jk}^\alpha=2\delta_{ij}\delta^{\alpha\beta},
\end{equation}
while the mass term $M_{ij}$ is a real antisymmetric matrix. We shall also assume that the single particle spectrum of $\hat H$ in (\ref{Dirac}) has a gap at zero energy, or equivalently that the matrix $\gamma_{ij}^\alpha k_\alpha - \rmi M_{ij}$ is invertible for all wave vectors $\mathbf{k} \in \R^d$. Let us call two mass terms $M_{ij}$ equivalent when the corresponding hamiltonians can be transformed one into the other without closing the gap. So the question arises: {\em how many inequivalent mass terms in (\ref{Dirac}) are possible?}.
\par
The above question should be refined by specifying the symmetry constraints imposed on the system. We shall not take into account any spatial symmetry apart of the obvious translational invariance of the Dirac hamiltonian (\ref{Dirac}). Instead, we shall consider the action of a compact group $G_{\rm F}$ in the Fock space over each spatial point $\mathbf{x}$. As follows from the discussion in sections \ref{sec:nambu} and \ref{sec:antiunitary}, the corresponding constraints on $H$ can be equivalently formulated in terms of the action of another group $G$ in the real Nambu space $W_\R$. This space can be conveniently identified with the mode space of Majorana fermions. Therefore, an element $g \in G$ represented in $W_\R$ by the matrix $g_{ij}$ acts on the mode space in the following way:
\begin{equation}
\label{local_g}
\hat c^i_\mathbf{x} \mapsto g_{ij} \hat c^j_{\mathbf{x}}.
\end{equation}
As explained in section \ref{sec:gaps}, the unitary symmetries form a subgroup $G_0 \subset G$, while the antiunitary or chiral symmetries, if present, correspond to its coset $G_1$.
\par
Let us now consider the isotypic decomposition of $W_\R$ given by equations (\ref{isotypic}) and (\ref{lambda_block}). Since the operators $M$ and $\gamma^\alpha$ defined by the matrices $M_{ij}$ and $\gamma^\alpha_{ij}$ are $G_0\mbox{-equivariant}$, they admit the following form:
\begin{equation}
\label{m_lambda}
\begin{aligned}
M=\bigoplus_{\lambda \in \hat G_0^{\rm r}} 1 \otimes_{F_\lambda} m_\lambda,\\
\gamma^\alpha=\bigoplus_{\lambda \in \hat G_0^{\rm r}} 1 \otimes_{F_\lambda} \gamma^\alpha_\lambda,
\end{aligned}
\end{equation}
where $m_\lambda$ and $\gamma^\alpha_\lambda$ act in the multiplicity space $E_\lambda$. It has been conjectured in \cite{kitaev} that the mass term can be continuously deformed to anticommute with $\gamma^\alpha$ without closing the gap. The same conjecture applies to each $m_\lambda$ individually, therefore such deformation can be performed while respecting all symmetry constraints. On the other hand, since the spectral flattening transformation (\ref{flatten}) preserves the anticommutation with $\gamma^\alpha_\lambda$, one can equivalently replace the mass term $M$ by
\begin{equation}
\label{tilde_M}
\tilde M=\bigoplus_{\lambda \in \hat G_0^{\rm r}} 1 \otimes_{F_\lambda} \tilde m_\lambda,
\end{equation}
where each $\tilde m_\lambda$ satisfies the following constraints:
\begin{equation}
\label{tilde_m}
\begin{aligned}
\tilde m_\lambda^2=-1\\
\tilde m_\lambda \gamma^\alpha_\lambda = - \gamma^\alpha_\lambda \tilde m_\lambda.
\end{aligned}
\end{equation}
Furthermore, if $G_1$ is not empty, analogously with equation (\ref{h_phi}) one has
\begin{equation}
\label{m_phi}
\tilde m_\lambda \phi_\lambda = - \phi_\lambda \tilde m_\lambda
\end{equation}
and
\begin{equation}
\label{gamma_phi}
\gamma^\alpha_\lambda \phi_\lambda = - \phi_\lambda \gamma^\alpha_\lambda.
\end{equation}
On the other hand, like $\tilde h$ in (\ref{h_f}), both $\gamma^\alpha_\lambda$ and $\tilde m_\lambda$ commute with the elements of $F_\lambda$:
\begin{equation}
\label{gamma_f}
\begin{aligned}
\gamma^\alpha_\lambda f=f \gamma^\alpha_\lambda\\
\tilde m_\lambda f = f \tilde m_\lambda,
\end{aligned}
\end{equation}
where $f \in F_\lambda$. Equations (\ref{gamma}), (\ref{tilde_m}), (\ref{gamma_phi}) and (\ref{gamma_f}) define the joint action of $\gamma^\alpha_\lambda$ and the Clifford algebras generated by $F_\lambda$ and $\phi_\lambda$ on $E_\lambda$. Thus, as has been pointed out in \cite{kitaev}, the Dirac matrices $\gamma^\alpha_\lambda$ extend the Clifford algebras $C^{p,q}$ or $C^n_\C$ in (\ref{extensions}) by adding $d$ positive or negative generators. The exact form of these generators is given in Tables  \ref{tab:real_dirac} and \ref{tab:complex_dirac}.
\begin{table}
\caption{\label{tab:real_dirac}Clifford module structure of $E_\lambda$ for the Dirac operator (\ref{Dirac}) in the real case. The element $\tilde m_\lambda$ is defined by equation (\ref{tilde_M}) and $\gamma^\alpha_\lambda$ by (\ref{m_lambda}). See also the caption of Table \ref{tab:real}. Note that $n$ corresponds here to the index of the symmetry class in Table \ref{tab:real}, and {\em not} to the $K\mbox{-theory}$ index of Clifford algebra extension.}
{\begin{tabular}{@{}*{5}{l}@{}}
\hline\\
&Cartan&Clifford algebra&Extending&Clifford algebra\cr
$n$&label&generators&generator&extension\cr
\hline\\
0&AI&$\{i,\,\phi_\lambda,\,\phi_\lambda\gamma^1_\lambda,\dots\phi_\lambda\gamma^d_\lambda\}$&$\phi_\lambda \tilde m_\lambda$&$C^{d+1,1} \to C^{d+1,2}$\cr
1&BDI&$\{\phi_\lambda,\,\gamma^1_\lambda,\dots\gamma^d_\lambda\}$&$\tilde m_\lambda$&$C^{0,d+1} \to C^{1,d+1}$\cr
2&D&$\{\gamma^1_\lambda,\dots\gamma^d_\lambda\}$&$\tilde m_\lambda$&$C^{0,d} \to C^{1,d}$\cr
3&DIII&$\{\phi_\lambda,\,\gamma^1_\lambda,\dots\gamma^d_\lambda\}$&$\tilde m_\lambda$&$C^{1,d} \to C^{2,d}$\cr
4&AII&$\{i,\,\phi_\lambda,\,\phi_\lambda\gamma^1_\lambda,\dots\phi_\lambda\gamma^d_\lambda\}$&$\phi_\lambda \tilde m_\lambda$&$C^{2,d} \to C^{3,d}$\cr
5&CII&$\{i,\,j,\,k\phi_\lambda,\,k\gamma^1_\lambda,\dots k\gamma^d_\lambda\}$&$k\tilde m_\lambda$&$C^{d+3,0} \to C^{d+3,1}$\cr
6&C&$\{i,\,j,\,k\gamma^1_\lambda,\dots k\gamma^d_\lambda\}$&$k\tilde m_\lambda$&$C^{d+2,0} \to C^{d+2,1}$\cr
7&CI&$\{i,\,j,\,k\phi_\lambda,\,k\gamma^1_\lambda,\dots k\gamma^d_\lambda\}$&$k\tilde m_\lambda$&$C^{d+2,1} \to C^{d+2,2}$\cr
\hline
\end{tabular}}
\end{table}
\par
\begin{table}
\caption{\label{tab:complex_dirac}Clifford module structure of for the Dirac operator (\ref{Dirac}) in the complex case. See also Table \ref{tab:real_dirac}.}
{\begin{tabular}{@{}*{5}{l}}
\hline\\
&Cartan&Clifford algebra&Extending&Clifford algebra\cr
$n$&label&generators&generator&extension\cr
\hline\\
0&A&$\{\gamma^1_\lambda,\dots\gamma^d_\lambda\}$&$\tilde m_\lambda$&$C_\C^{d} \to C_\C^{d+1}$\cr
1&AIII&$\{\phi_\lambda,\,\gamma^1_\lambda,\dots\gamma^d_\lambda\}$&$\tilde m_\lambda$&$C_\C^{d+1} \to C_\C^{d+2}$\cr
\hline
\end{tabular}}
\end{table}
\par
As follows from these tables, the isotypic component $\tilde m_\lambda$ of the mass term $M$ extends the action of the Clifford algebra generated by $F_\lambda$, $\phi_\lambda$ and $\gamma^\alpha_\lambda$ in the following way:
\begin{equation}
\label{dirac_extensions}
\begin{aligned}
C^{p,q+d}&\to C^{p+1, q+d};\\
C^{p+d,q}&\to C^{p+d, q+1};\\
C^{n+d}_\C&\to C^{n+d+1}_\C.
\end{aligned}
\end{equation}
Equations (\ref{n_real}) show that in the real case the extensions with equal values of $n-d \bmod 8$ are Morita equivalent, as are the extensions with equal values of $n-d \bmod 2$ in the complex case. This phenomenon is the origin of the periodicity in properties of topological insulators with respect to the spatial dimension \cite{kitaev}.
\par
The space of allowed operators $\tilde m_\lambda$ can be characterized by the same methods as the spaces $\tilde{\cal S}_\lambda$ in section \ref{sec:gaps}, yielding results similar to equations (\ref{sum_to_simple}), (\ref{simple_to_sum}) and (\ref{simple_to_simple}) up to a dimension shift. However, we are now interested only in the number ${\cal N}$ of possible inequivalent mass terms in (\ref{Dirac}). This number is given by the formula
$$
{\cal N}=\prod_{\lambda \in G_0^{\rm ra}} {\cal N}_\lambda,
$$
where ${\cal N}_\lambda$ is the number of connected components in the classifying space of $\tilde m_\lambda$. Depending upon the value of $n-d$, the following three situations are possible:
\begin{enumerate}
\item\label{case:m_1}{${\cal N}_\lambda=1$. All $\tilde m_\lambda$ are equivalent. This occurs for $n-d =3,\,5,\,6,\mbox{ or }7 \bmod 8$ in the real case and for $n-d=1 \bmod 2$ in the complex case.}
\item\label{case:m_2}{${\cal N}_\lambda=2$. This is the case when the space of allowed $\tilde m_\lambda$ is either $\mathrm{O}(k)$ or $\mathrm{O}(2k)/\mathrm{U}(k)$. This occurs only in the real case when $n-d =1\mbox{ or }2 \bmod 8$.}
\item\label{case:m_k1}{${\cal N}_\lambda=k+1$. The space of $\tilde m_\lambda$ is a union of a $k+1$ Grassmannian manifolds similar to that of (\ref{simple_to_sum}) and the value of $k$ depends on the dimension of the multiplicity space $E_\lambda$. This occurs for $n-d =0,\mbox{ or }4 \bmod 8$ in the real case and for $n-d=0 \bmod 2$ in the complex case.}
\end{enumerate}
\par
It is instructive to consider the operation of direct sum of the fermion mode spaces. This operation equips the set of equivalence classes of $\tilde m_\lambda$ with the structure of a semi-group, which is isomorphic to the trivial group, to $\Z_2$ or to the semi-group $\N$ of non-negative integers in the cases \ref{case:m_1}, \ref{case:m_2} or \ref{case:m_k1} respectively. Therefore, the cases  \ref{case:m_k1} and \ref{case:m_2} are distinct even if $k=1$.
\par
One possible way to study the equivalence classes of $M$ is provided by $K\mbox{-theory}$. This approach consists in considering formal differences between these classes and effectively working in infinite dimensional multiplicity spaces, which correspond to $k\to\infty$ in case \ref{case:m_k1}. However, equation (\ref{Dirac}) can also be used to describe the physical systems where this limit does not make sense and where the value of ${\cal N}_\lambda=k+1$ remains always finite. This value depends on $n$, $d$ and the dimension of $E_\lambda$. Let us illustrate this point by the case of class CI ($n=7$, $F_\lambda=\HH$) in three dimensions. As follows from Table \ref{tab:real_dirac} this case corresponds to the Clifford algebra extension $C^{5,1} \to C^{5,2}$. Since the algebras $C^{5,1}=\HH(4)$ and $C^{5,2}=\HH(4)\oplus \HH(4)$ act on the quaternionic vector space $E_\lambda$, the dimension of $E_\lambda$ is a multiple of 4, and the value of $k$ is given by the formula
$$
k=\frac{\mathrm{dim}_{\HH}(E_\lambda)}{4}.
$$
The relations between $k$ and $\mathrm{dim}_{F_\lambda}(E_\lambda)$ for other combinations of $n$ and $d$ can be obtained in a similar manner.
\section{Summary and discussion}\label{sec:discussion}
In the present paper we have proposed an approach to the classification of topological phases of insulators and superconductors, synthesizing the methodology developed in \cite{kitaev} and \cite{hhz, zirnbauer2010}. The main object in this study is the set of meta-symmetries remaining after elimination of ordinary symmetry constraints, following the approach proposed in \cite{hhz}. It is shown, that these meta-symmetries provide the multiplicity spaces of each isotypic component of the real Nambu space $W_\R$ with the structure of a Clifford module. The ten symmetry classes of fermion hamiltonians described in  \cite{hhz} are in a one-to-one correspondence with the ten Morita equivalence classes of the real and complex Clifford algebras. When the single-particle spectrum of the hamiltonian contains a gap at zero energy, the action of the Clifford algebra can be extended by adding a new generator, as described in \cite{kitaev}. This leads to the complete classification of the homotopy classes of the gapped fermion hamiltonians in the case of a compact symmetry group.
\par
The main difference of our approach from that of \cite{hhz} consists in working with the isotypic decomposition of the Nambu space over the real numbers. This allows one to take into account the natural real structure of the Nambu representation directly, instead of treating it as an externally imposed condition. As a result, the hamiltonian acts independently on each isotypic component of $W_\R$, simplifying matters compared to the construction used in \cite{hhz}. The underlying complexity is thus encoded in the Clifford module structure of the multiplicity space $E_\lambda$. As concerns the antiunitary symmetries, we generalize the approach of \cite{hhz} by considering all antiunitary operators in the Fock space which are compatible with the quadratic hamiltonians. It appears that both antiunitary and chiral symmetries give rise to equivalent symmetry constraints in the Nambu space representation.
\par
The classification of the topological insulators and superconductors based on Clifford modules has been first proposed in \cite{kitaev}, where explicit formulas were given for three out of ten Altland-Zirnbauer symmetry classes (D, DIII and AII). The present paper completes this classification. There is an important distinction, however, in our interpretation of the class AII. The operator $Q$, which is proposed as one of the generators of the Clifford algebra $C^{2,0}$ is treated in \cite{kitaev} as a generator of the charge conservation symmetry $\mathrm{U}(1)$, whereas we interpret it as a member of the algebra of intertwining operators $F_\lambda=\C$, corresponding to the imaginary unit of $\C$.
\par
The relationship between Altland-Zirnbauer symmetry classes, topological insulators and Clifford algebras has also been considered in \cite{stone}. The starting point of this approach consists in imposing the action of $C^{p,0}$ on the space of states. An observation is then made that the set of possible extensions of this representation to $C^{p+1, 0}$ is a symmetric space or a union of symmetric spaces, which are identified with the symmetric spaces of the time evolution operators in the classification by \cite{az}. The hamiltonian and the discrete antiunitary symmetries are then constructed explicitly from the generators of $C^{p,0}$ to satisfy the imposed commutation and anticommutation conditions. The main difference of this approach with ours is that we do not treat the action of the Clifford algebra as an ordinary symmetry which should be imposed from the beginning, but as a meta-symmetry, as explained above. As a result, the Clifford module structure emerges naturally, which guarantees the exhaustiveness of the classification.
\par
Whilst $K\mbox{-theory}$ is a powerful mathematical tool, its application to the problem under consideration has some drawbacks. Indeed, since $K\mbox{-theory}$ works with the formal differences of the Clifford modules instead of the modules of themselves, it characterizes only the stable homotopy properties. This is equivalent to considering physical systems up to an augmentation (adding extra dimensions to the state space). Though such augmentation, as pointed out in \cite{kitaev}, simplifies the topological classification, the dimension of the state space of real physical systems is always finite (and may be small). Therefore, a complete homotopy classification of gapped hamiltonians is a problem of interest on its own.
\par
It is instructive to compare our approach to the classification of topological phases proposed in  \cite{schnyder}. The latter is based on the consideration of two antiunitary operations --- time reversal symmetry (TRS) and particle-hole conjugation (PHS). Whereas the PHS is interpreted in \cite{schnyder} as a physical symmetry on its own, it corresponds in our approach to the composition of $C$ defined in (\ref{C}) and an intertwining operator of unit norm from $F_\lambda$. In the case $F_\lambda=\C$ or $\HH$, the imaginary unit $i \in F_\lambda$ can be identified with the operator $J$ defined in (\ref{J}). Since $J$ anticommutes with the particle-hole symmetry, the latter cannot occur for $F_\lambda=\C$, and in the case $F_\lambda=\HH$ the corresponding operator squares to $-1$. Therefore, the cases denoted symbolically as $\mbox{PHS}=+1$, $0$ or $-1$ in \cite{schnyder} are in one to one correspondence with $F_\lambda=\R$, $\C$ or $\HH$ in our interpretation.
\par
The situation with the time reversal symmetry is more complicated. First of all, one should distinguish between the time reversal operator $T$ in the Nambu space $W$ and that in $E_\lambda$. This point is explained in greater details in \cite{hhz, zirnbauer2010}, where the corresponding operator in $E_\lambda$ is called the ``transferred time reversal''. The symmetry class is determined by the sign of the square of the transferred time reversal, which might be different from the sign of $T^2$ (this may occur for classes AI and AII, as explained in section \ref{sec:antiunitary}). The comparison of Tables \ref{tab:real} and \ref{tab:complex} with Table II of \cite{schnyder} shows that in our approach $\phi_\lambda$ behaves as the transferred time reversal operator, except for classes CI and CII. To understand this discrepancy, let us recall, that $\phi_\lambda$ acts in the left $F_\lambda\mbox{-module}$ $E_\lambda$ and is defined by the condition of $F_\lambda\mbox{-linearity}$ (or antilinearity in the case $F_\lambda=\C$). In contrast to that, the transferred time reversal introduced in \cite{hhz, zirnbauer2010} is an antiunitary operator acting in a complex linear space. In our approach it should correspond to $f \phi_\lambda$ for some scalar of unit norm $f\in F_\lambda$. In the case $F_\lambda=\HH$ the antiunitarity of the transferred time reversal implies that it anticommutes with the operator $J$ in (\ref{J}). After identifying $J$ with the action of $i\in \HH$ as above, one get $if=-fi$. Therefore $f=\cos(\alpha) j + \sin(\alpha) k$ for some real $\alpha$, yielding $f^2=-1$. Hence, in the case of classes CI and CII the transferred time reversal operator squares to the opposite of $\phi_\lambda^2$, in consistence with Table II of \cite{schnyder}. Note finally that $\phi_\lambda$ is defined as long as the set $G_1$ of antiunitary or chiral symmetries is not empty, while the time reversal symmetry is not necessarily present (that is the case when none of the antiunitary operators commute with the action of $G_0$ \cite{dyson}).
\par
After this work was completed, we became aware of Ref. \cite{fidkowski}, where, among other issues the authors proposed a novel approach for the classification of Altland-Zirnbauer symmetry classes. They suggested to consider $\R G$ as a semi-simple {\em graded} algebra with the even subalgebra  $\R G_0$. The action of $\R G$ on the isotypic component ${\cal W}_\lambda$ (or ${\cal W}_\lambda \oplus {\cal W}_{\lambda'}$ when blocks ${\cal W}_\lambda$ and ${\cal W}_{\lambda'}$ are permuted by the action of odd elements of $\R G$) corresponds to one of the simple components of $\R G$ and as such can be classified by the general theory developed by Wall \cite{wall}. The advantage of this approach is that the symmetry classes are further structured by identifying them with the elements of the graded Brauer group of the field with respect to which the simple component of $\R G$ is central. This group is isomorphic to $\Z_8$ for $\R$ and to $Z_2$ for $\C$.
\section*{Acknowledgments}
The authors are grateful for the stimulating
discussions to M.~Rovinski, E.~Amerik, M.~Verbitsky, P.~Delplace, D.~Sticlet, J.-N.~Fuchs, F.~Piechon and
P.~Pushkar'.
\appendix
\section*{Appendix}
\setcounter{section}{1}
This appendix contains the proof of the Corollaries 1 and 2.
\begin{proof}[Proof of the Corollary 1]
The definition (\ref{f_action}) of the left action of $F_\lambda$ on $E_\lambda$ implies that the evaluation map (\ref{evaluation}) is middle-linear \cite{hungerford}:
$$
\epsilon_\lambda(rf, e)=\epsilon_\lambda(r, fe)
$$
and hence it factors through the tensor product (\ref{lambda_block}):
$$
\xymatrix{{\cal R}_\lambda \times E_\lambda \ar[dr]\ar[rr]^{\epsilon_\lambda} &&W_\R\\
& {\cal R}_\lambda \otimes_{F_\lambda} E_\lambda \ar[ur]^{\iota_\lambda}}
$$
Since ${\cal W}_\lambda$ is by definition the image of $\epsilon_\lambda$, the $\R G_0\mbox{-module}$ homomorphism $\iota_\lambda$ maps ${\cal R}_\lambda \otimes_{F_\lambda} E_\lambda$ onto ${\cal W}_\lambda$. It remains to prove the injectivity of $\iota_\lambda$. Let us assume the converse and denote the kernel of $\iota_\lambda$ by $K$. Consider the following exact sequence of $\R G_0\mbox{-modules}$:
$$
\xymatrix{0\ar[r]&K\ar[r]^(0.33){\mu} &{\cal R}_\lambda \otimes_{F_\lambda} E_\lambda \ar[r]^(0.67){\iota_\lambda}&W_\R}
$$
and apply the functor $\Hom_{G_0}({\cal R}_\lambda, -)$ to it. Since $F_\lambda$ is a division algebra, $E_\lambda$ is a free $F_\lambda\mbox{-module}$ \cite{hungerford}, and we get the exact sequence
$$
\xymatrix{0\ar[r]&\Hom_{G_0}({\cal R}_\lambda,K)\ar[r]^(0.4){\mu_*} &\Hom_{G_0}({\cal R}_\lambda, {\cal R}_\lambda)\otimes_{F_\lambda}E_\lambda \ar[r]^(0.75){\iota_{\lambda *}}& E_\lambda}.
$$
By construction, $\iota_{\lambda *}(\mathrm{id}_{{\cal R}_\lambda} \otimes_{F_\lambda} e)=e$, and since $\Hom_{G_0}({\cal R}_\lambda, {\cal R}_\lambda)=F_\lambda$, the map $\iota_{\lambda *}$ is an isomorphism. Therefore
\begin{equation}
\label{no_kernel}
\Hom_{G_0}({\cal R}_\lambda,K)=0.
\end{equation}
If $K$ is not empty, it contains at least one irreducible representation of $G_0$. Denote the corresponding subspace by $L$. Let $\{e_i\}$ be the basis of $E_\lambda$ as an $F_\lambda\mbox{-module}$ and let $\pi_i: E_\lambda \to F_\lambda$ be the corresponding projections. Then, there exists $\pi_j$ such that $(1\otimes_{F_\lambda}\pi_j)\circ\mu(L)\neq 0$. By virtue of the Schur lemma, $(1\otimes_{F_\lambda}\pi_j)\circ\mu$ establishes a $G_0\mbox{-equivariant}$ isomorphism between $L$ and ${\cal R}_\lambda$, which contradicts (\ref{no_kernel}). Therefore, $\iota_\lambda$ is injective.
\end{proof}
\begin{proof}[Proof of the Corollary 2]
Let $\sigma$ be an automorphism of $G_0$ defined by
\begin{equation}
\label{sigma}
\sigma(g)= sgs^{-1}.
\end{equation}
The action of $\sigma(g)$ on ${\cal R}_\lambda$ is a real representation of $G_0$. This representation is obviously irreducible and hence corresponds to some type $\lambda' \in \hat G_0^{\rm r}$. The algebras of intertwining operators $F_\lambda$ and $F_{\lambda'}$ are clearly isomorphic. Since the action of $s$ transforms any submodule of $W_\R$ isomorphic to ${\cal R}_\lambda$ into one isomorphic to ${\cal R}_{\lambda'}$, one has
\begin{equation}
\label{s_include}
s({\cal W}_\lambda) \subset {\cal W}_{\lambda'}.
\end{equation}
On the other hand $s^2\in G_0$, therefore we have
\begin{equation}
\label{s_square}
s^2({\cal W}_\lambda) = {\cal W}_\lambda.
\end{equation}
The relations (\ref{s1}) and (\ref{s2}) follow from (\ref{s_include}) and (\ref{s_square}).
\par
Let $\{e_i\}$ and $\{e_i'\}$  be the bases of the free modules $E_\lambda$ and $E_{\lambda'}$. Consider the action of $s$ on $r\otimes_{F_\lambda}e_i$ for a given $r \in {\cal R}_\lambda$:
\begin{equation}
\label{s_w}
s\left(r\otimes_{F_\lambda}e_i \right)= \sum_j \eta_{ij}(r) \otimes_{F_\lambda}e_j'.
\end{equation}
The above formula defines $\R\mbox{-linear}$ maps
$$
\eta_{ij}: {\cal R}_\lambda \to {\cal R}_{\lambda'}.
$$
Then the following diagram is commutative
\begin{equation}
\label{square}
\xymatrix{{\cal R}_\lambda\ar[r]^{\eta_{ij}}\ar[d]^g &{\cal R}_{\lambda'}\ar[d]^{\sigma(g)}\\
{\cal R}_\lambda\ar[r]^{\eta_{ij}} &{\cal R}_{\lambda'}}
\end{equation}
Here the map $g$ is the action of the group element $g\in G_0$ and $\sigma$ is defined in (\ref{sigma}).
Select an arbitrary non-zero map $\eta_{ij}$ and call it $\xi_{s, \lambda}$. Since the representation ${\cal R}_{\lambda'}$ is irreducible, $\xi_{s, \lambda}$ is bijective. Consider the following commutative diagram:
$$
\xymatrix{{\cal R}_\lambda\ar[r]^{\eta_{ij}}\ar[d]^g &{\cal R}_{\lambda'}\ar[d]^{\sigma(g)}\ar[r]^{\xi_{s, \lambda}^{-1}}& {\cal R}_\lambda \ar[d]^g \\
{\cal R}_\lambda\ar[r]^{\eta_{ij}} &{\cal R}_{\lambda'}\ar[r]^{\xi_{s, \lambda}^{-1}}& {\cal R}_\lambda}
$$
We see that the map $\xi_{s, \lambda}^{-1} \circ \eta_{ij}$ intertwines the action of $G_0$ on ${\cal R}_\lambda$. Hence there exist $\psi_{ij} \in F_{\lambda}$ such that
$$
\eta_{ij}(r)=\xi_{s, \lambda}(r) \psi_{ij}.
$$
One might be tempted to define the map
$$
\phi_\lambda: E_\lambda \to E_{\lambda'}
$$
by the condition
\begin{equation}
\label{phi_basis}
\phi_\lambda(e_i)=\sum_j \psi_{ij} e_j'
\end{equation}
and use the middle-linearity of the tensor product to represent the action (\ref{s_w}) of $s$ on ${\cal W}_\lambda$ as follows:
\begin{equation}
\label{pure}
s \Bigr|_{{\cal W}_\lambda}=\xi_{s, \lambda} \otimes_{F_\lambda} \phi_\lambda.
\end{equation}
However, we must proceed here with caution. Indeed, if $\xi_{s, \lambda}$ is not $F_\lambda\mbox{-linear}$, so must not be $\phi_\lambda$ in order for the tensor product in (\ref{pure}) to be well defined. One can not thus use $F_\lambda\mbox{-linearity}$ to extend the definition (\ref{phi_basis}) from the basis vectors to the entire space $E_\lambda$. Let us examine the situation more closely. Define the map $f': {\cal R}_{\lambda'} \to {\cal R}_{\lambda'}$ by $f'=\xi_{s, \lambda} \circ f \circ \xi_{s, \lambda}^{-1}$ (here we use the same symbol $f$ to denote an element $f\in F_\lambda$ and its action on ${\cal R}_{\lambda}$ by the the right multiplication). Then the top and the bottom squares of the following cube diagram are commutative:
$$
\begin{array}{c}
\xymatrix@R=2pc@C=2pc@L=0pt{
{\cal R}_\lambda \ar[rr]^*+{\xi_{s, \lambda}} \ar[dr]^(0.6)*+{f}
\ar[dd]_*+{g} && {\cal R}_{\lambda'} \ar[dr]^*+{f'} \ar[dd]|\hole^(0.75)*+{\sigma(g)} & \\
& {\cal R}_\lambda \ar[dd]_(0.25)*+{g} \ar[rr]^(0.25)*+{\xi_{s, \lambda}} &&
{\cal R}_{\lambda'} \ar[dd]^*+{\sigma(g)} \\
{\cal R}_\lambda \ar[rr]|\hole_(0.7)*+{\xi_{s, \lambda}} \ar[dr]_*+{f} &&
{\cal R}_{\lambda'} \ar[dr]^(0.35)*+{f'} & \\
& {\cal R}_\lambda \ar[rr]_*+{\xi_{s, \lambda}} & & {\cal R}_{\lambda'} }
\end{array}
$$
As we have already seen in (\ref{square}), the front and the back squares are commutative. The left square is commutative by the definition of $f$. It then follows that the right square is also commutative. Hence $f'$ intertwines the action of $G_0$ on ${\cal R}_{\lambda'}$ and corresponds to the right multiplication by some other element $f' \in F_\lambda$. This defines a map
\begin{eqnarray}
\label{beta}
\beta_\lambda: F_\lambda \to F_\lambda \nonumber \\
\beta_\lambda(f) = f' ,
\end{eqnarray}
which is clearly an $\R\mbox{-linear}$ automorphism of $F_\lambda$. We can now define $\phi_\lambda$ on the entire space $E_\lambda$ as follows
\begin{equation}
\label{phi}
\phi_\lambda\left(\sum_i f_i e_i \right)=\sum_{i, j} \beta_\lambda (f_i) \psi_{ij} e_j' .
\end{equation}
One can easily check that formula (\ref{pure}) with $\phi_\lambda$ given by (\ref{phi}) is consistent with the tensor product structure of both ${\cal R}_\lambda \otimes_{F_\lambda}E_\lambda$ and ${\cal R}_{\lambda'}\otimes_{F_\lambda}E_{\lambda'}$. As a final remark, we point out that, since the formula (\ref{pure}) can be extended to the entire coset $G_1$ by setting
\begin{equation}
\label{pure_coset}
sg \Bigr|_{{\cal W}_\lambda}=\left(\xi_{s, \lambda} \circ g \right) \otimes_{F_\lambda} \phi_\lambda,
\end{equation}
$\phi_\lambda$ and $\beta_\lambda$ do not depend on the choice of $s$.
\par
Let us now discuss the cases of $F_\lambda=\R$, $\C$ and $\HH$ separately. First of all, consider the situations where $\beta_\lambda$ can be chosen to be trivial. This is obviously the case for $F_\lambda=\R$. In the algebra of quaternions, all automorphisms of $\HH$ are internal \cite{hazewinkel},  that is one has $\beta_\lambda(f)=qfq^{-1}$ for some $q\in\HH$. We can thus replace $\xi_{s, \lambda}$ and $\phi_\lambda$ in (\ref{pure})  in the following way:
\begin{eqnarray}
\xi_{s, \lambda} \mapsto \xi_{s, \lambda} \, q \nonumber \\
\phi_\lambda \mapsto q^{-1} \phi_\lambda , \nonumber
\end{eqnarray}
and get a trivial automorphism $\beta_\lambda(f)=f$. In the remaining case $F_\lambda=\C$ there are two possible $\R\mbox{-linear}$ automorphisms: $\beta_\lambda(f)=f$ and $\beta_\lambda(f)=\bar f$. Thus the maps $\xi_{s, \lambda}$ and $\phi_\lambda$ can always be made $F_\lambda\mbox{-linear}$ except for the case $F_\lambda=\C$ where they may be $\C\mbox{-antilinear}$.
\par
Let us now consider specifically the case $\lambda=\lambda'$. Since the element $g=s^2$ belongs to $G_0$, the following holds:
$$
\xi_{s, \lambda}^2 \otimes_{F_\lambda} \phi_\lambda^2 = g \otimes_{F_\lambda} 1.
$$
This implies that $\phi_\lambda^2$ acts on $E_\lambda$ by left multiplication by some non-zero scalar $f_\phi \in F_\lambda$. On the other hand, $\phi_\lambda^2$ is always $F_\lambda\mbox{-linear}$, hence $f_\phi$ must be real for $F_\lambda=\HH$. Thus, in the cases $F_\lambda=\R$ or $F_\lambda=\HH$ an appropriate scaling of $\phi_\lambda$ makes $\phi_\lambda^2$ equal to either $+1_{E_\lambda}$ or $-1_{E_\lambda}$. Consider now the case $F_\lambda=\C$. Since
$$
f_\phi \phi_\lambda =(\phi_\lambda^2)\cdot\phi_\lambda=\phi_\lambda\cdot(\phi_\lambda^2)=\beta(f_\phi)\phi_\lambda,
$$
one has
$$
\beta_\lambda(f_\phi)=f_\phi .
$$
In the case $\beta_\lambda(f)=\bar f$ this implies $f_\phi \in \R$, hence $\phi_\lambda^2$ once again can be rescaled to $+1_{E_\lambda}$ or $-1_{E_\lambda}$. It remains to consider the situation where $F_\lambda=\C$ and both $\xi_{s, \lambda}$ and $\phi_\lambda$ are $\C\mbox{-linear}$. In this case it is possible to rescale $\phi_\lambda$ by a complex factor. Thus, one can always choose $\phi_\lambda^2=+1_{E_\lambda}$.
\end{proof}
\setcounter{section}{1}

\end{document}